\documentclass[12pt,titlepage,a4paper]{article}

\usepackage{amssymb}

\advance\textwidth1.5cm
\advance\textheight1cm
\advance\oddsidemargin-1.2cm
\advance\topmargin-1cm

\newcommand{\be}{\begin{equation}}
\newcommand{\ee}{\end{equation}}
\newcommand{\bea}{\begin{eqnarray}}
\newcommand{\eea}{\end{eqnarray}}
\newcommand{\beas}{\begin{eqnarray*}}
\newcommand{\eeas}{\end{eqnarray*}}

\newcommand{\bs}{\begin{sloppypar}}
\newcommand{\es}{\end{sloppypar}}

\def\id{{\mathchoice
{\rm 1 \mskip-4mu l}
{\rm 1 \mskip-4mu l}
{\rm 1 \mskip-4.5mu l}
{\rm 1 \mskip-5mu l}}}

\newcommand{\B}{{\cal B}}
\newcommand{\J}{{\cal J}}
\newcommand{\JT}{\tilde{\cal J}}
\renewcommand{\O}{{\cal O}}
\renewcommand{\P}{{\cal P}}
\newcommand{\Q}{{\cal Q}}
\newcommand{\R}{{\cal R}}
\renewcommand{\S}{{\cal S}}
\newcommand{\T}{{\cal T}}
\newcommand{\Z}{{\cal Z}}

\newcommand{\JK}{{\bf J}}
\newcommand{\JKT}{\tilde{\JK}}

\newcommand{\BH}{\widehat{\cal B}}
\newcommand{\JH}{\widehat{\cal J}}

\newcommand{\PH}{\widehat{\cal P}}

\newcommand{\SH}{\widehat{\cal S}}
\renewcommand{\TH}{\widehat{\cal T}}

\newcommand{\F}{{\mathfrak F}}
\newcommand{\I}{{\mathfrak I}}
\newcommand{\BI}{\bar{\mathfrak I}}
\newcommand{\U}{{\mathfrak U}}
\newcommand{\V}{{\mathfrak V}}

\newcommand{\FH}{\widehat{\mathfrak F}}
\newcommand{\IH}{\widehat{\mathfrak I}}

\newcommand{\UH}{\widehat{\mathfrak U}}

\newcommand{\SJ}{{\sf J}}
\renewcommand{\SS}{{\sf S}}
\newcommand{\ST}{{\sf T}}

\newcommand{\SJH}{\widehat{\sf J}}
\newcommand{\SSH}{\widehat{\sf S}}
\newcommand{\STH}{\widehat{\sf T}}

\newcommand{\HJ}{\widehat{J}}
\newcommand{\HS}{\widehat{S}}
\newcommand{\HT}{\widehat{T}}

\newcommand{\JHQ}{\JH \!\!\!\! \backslash}
\newcommand{\BHQ}{\BH \!\!\!   \backslash}
\newcommand{\SHQ}{\SH \!\!\!   \backslash}
\newcommand{\THQ}{\TH \!\!\!\! \backslash}

\newcommand{\HJQ}{\HJ \!\!\!   \backslash}

\newcommand{\HSQ}{\HS \!\!\!   \backslash}
\newcommand{\HTQ}{\HT \!\!\!\! \backslash}

\newcommand{\ga}{{\mathfrak a}}
\newcommand{\gf}{{\mathfrak f}}
\renewcommand{\gg}{{\mathfrak g}}
\newcommand{\gh}{{\mathfrak h}}
\newcommand{\gi}{{\mathfrak i}}
\newcommand{\bgi}{\bar{\mathfrak i}}
\newcommand{\gm}{{\mathfrak m}}

\newcommand{\gah}{\widehat{\mathfrak a}}

\newcommand{\ggh}{\widehat{\mathfrak g}}
\newcommand{\ghh}{\widehat{\mathfrak h}}

\newcommand{\N}{\mathbb{N}}
\newcommand{\C}{\mathbb{C}}
\newcommand{\Zet}{\mathbb{Z}}

\newcommand{\ap}{\alpha^\prime}
\renewcommand{\d}{{\rm d}}
\newcommand{\p}{\prime}

\newcommand{\aoo}{algebra of observables}
\renewcommand{\a}{algebra}
\renewcommand{\c}{constraint}
\newcommand{\obs}{observable}

\renewcommand{\r}{representation}
\newcommand{\ir}{irreducible}
\newcommand{\ie}{{\it i.e.}}
\newcommand{\eg}{{\it e.g.}}

\begin{document}

\begin{titlepage}

\vspace*{-2cm}
\begin{flushright} { THEP 98/8\\University of Freiburg\\April 1998\\
                     hep-th/9805057}
\end{flushright}
\vspace{2cm}

\begin{center}
{\large\bf The Nambu--Goto Theory of Closed Bosonic Strings\\[6mm]
           Moving in $1+3$--Dimensional Minkowski Space: \\[7mm]
           The Quantum Algebra of Observables}\\[2.5cm]
{ K. Pohlmeyer}\\[4mm]
{Fakult\"at f\"ur Physik der Universit\"at Freiburg,
Hermann--Herder--Str.\ 3, \\[3mm]
D--79104 Freiburg,
Germany}\\[4cm]
{\bf Abstract}\\[7mm]
\end{center}

\noindent
A relevant part of the quantum \aoo\ for the closed bosonic
strings moving in $1+3$--dimensional Minkowski space is
presented in the form of generating relations involving
still one, as yet undetermined, real free parameter.

\end{titlepage}

\section*{Introduction}

The present communication is part of an ongoing effort to describe
the observable features of the Nambu--Goto field theory of relativistic,
linearly extended geometric objects, called strings, in purely algebraic
terms \cite{Uncov}.

Conventionally, the Nambu--Goto theory is
treated as any field theory of point-like objects with phase space
\c s: auxiliary fields are 
introduced in order to give the action a quadratic form, BRST techniques
are used to handle the \c s, special
coordinates are introduced in order to ``solve'' the equations of
motion, to define new dynamical variables and to facilitate the 
transition from the classical theory to the quantum theory, etc.
({\it cf}, for instance, Ref.\ \cite{Brink}).
It is well-known that in its Euclidean version the classical
Nambu--Goto theory of closed strings coincides with the theory
of extremal surfaces (in whatever ambient space) and the global 
features of such extremal surfaces are by no means a trivial issue.

In the approach advocated here the string is not resolved into
its local pieces. For closed bosonic strings moving in $1+(d-1)
$--dimensional Minkowski space -- in the following I shall
restrict myself to this case -- the differential geometric
information (typically within a finite region of time and space) 
about the trajectory surface of the string is encoded in a countable
set of piecewise conserved data independent of any chosen
coordinatization. These observable data are called invariant charges.
Since they are well enough localized and flexibly localized
in ambient time and space, the computation of their mutual
Poisson brackets does not pose any difficulties nor does it
produce any ambiguities. Moreover, the polynomial ring of 
invariant charges closes under Poisson bracket operation.
Hence this ring forms a Poisson algebra. The next task is 
a presentation of this algebra. It would be overly naive
to expect a presentation in terms of classical Lie algebras.
Instead one is confronted with the typical situation in
{\it combinatorial algebra}: a description of the pertinent
algebra in terms of relations imposed on a freely generated algebra
or, more technically speaking, a description in terms of a
quotient of a freely generated algebra by an ideal, the latter being 
generated by the relations just mentioned \cite{Stan}. The difficulty
with this description
lies in the lack of an explicit characterization of those elements
of the freely generated algebra which are contained in the ideal,
or even in the lack of control over the ``dimension'' of the
ideal. Unfortunately, no alternative ways of presenting 
the Poisson algebra under consideration are available. (Obviously,
for the matter of higher dimensional extended geometric objects,
the situation is not going to improve!) However, once a sufficiently
large and coherent class of generators for the (classical) ideal
has been identified, the passage to a relevant part of the associative 
quantum \aoo\
is achieved by means of a deformation of the generators based upon
correspondence and consistency considerations. It is the central
concern of the following sections to demonstrate for strings
moving in $1+3$--dimensional Minkowski space how this deformation 
works in detail and which classical preparations are required.

In the past, the efforts of my collaborators and myself aimed at
a presentation of the \aoo\ for strings moving in $1+2$--dimensional
Minkowski space. The reason for choosing these particular dimensions
was the simple structure of the stabilizer group $SO(2)$ of the 
momentum rest frame for the string. A ``slight'' irregu- larity
concerning the assignment of structural roles to the various generators
of the algebra was the price for the technical simplifications.
In the meantime another (possibly related) drawback made itself 
felt: the observed systematics of the generators for the aforementioned
ideal suggests that the final generators to appear
involve among a vast variety of 
terms eightfold Poisson brackets for the classical theory and eightfold
commutators for the quantum theory, respectively.
The handling and processing of such expressions is beyond the
computational means of my collaborators and myself. By contrast,
for strings moving in $1+3$--dimensional Minkowski space there are 
indications that the collection of generators will be complete at a
much earlier stage. This makes it worthwhile to pass to $1+3$--dimensional
ambient Minkowski space and to put up with the complications reflected
in the employment of Clebsch--Gordan coefficients and $6j$--symbols.

\section*{Classical Preparations}

Consider the algebra formed by the generators of the infinitesimal
\obs\ symmetry transformations of the Nambu--Goto theory of closed
bosonic strings moving in $1+3$--dimensional Minkowski space. The
structure of this algebra is given in terms of the Poincar\'e
algebra corresponding to rigid Lorentz transformations and
translations of the string in the ambient Minkowski space and of
the algebra of (internal) invariant charges $\Z^+_{\mu_1\dots\mu
_N}$, built from the left movers
\[     u^+_{\mu_i}(\tau,\sigma) := p_{\mu_i}(\tau,\sigma) +
    \frac1{2\pi\ap} \: \partial_\sigma x_{\mu_i}(\tau,\sigma) , \]
and $\Z^-_{\mu_1\dots\mu_N}$, built from the right movers
\[     u^-_{\mu_i}(\tau,\sigma) := p_{\mu_i}(\tau,\sigma) -
    \frac1{2\pi\ap} \: \partial_\sigma x_{\mu_i}(\tau,\sigma) . \]
Here $x_{\mu_i}(\tau,\sigma)$ and $p_{\mu_i}(\tau,\sigma)$ are the
canonical string variables as functions of a ``timelike'' parameter
$\tau$ and a ``spacelike'' parameter $\sigma$. The symbol $\ap$
denotes the inverse string tension. $\mu_i \in \{ 0,\dots,3 \}$,
$i = 1,\dots,N$, $N=1,2,3,\dots\:$.
\[  \Z^\pm_{\mu_1\dots\mu_N} := \oint \d\sigma \; u^\pm_{\mu_1}
    (\tau,\sigma) \: \R^\pm_{\mu_2\dots\mu_N}(\tau,\sigma) ,  \]
\[  \R^\pm_{\mu_2\dots\mu_N}(\tau,\sigma) :=
    \!\!\!\!\!\!\!\!\!\!\!\!\!\!\!\!\!\!\!\!\! \int\limits_{\qquad
    \qquad \sigma+\omega(\tau) > \sigma_2 > \sigma_3 > \dots > \sigma_N
    > \sigma} \!\!\!\!\!\!\!\!\!\!\!\!\!\!\!\!\!\!\!\!\!\!\!\!\!\!\!
    \!\!\!\!\!\!\!\!  \d\sigma_2  \cdots \int \d\sigma_N \quad
    \prod_{i=2}^N u^\pm_{\mu_i}(\tau,\sigma_i) .\]
In the last formula the symbol $\omega(\tau)$ stands for the period
of the canonical string variables as functions of the (closed string)
curve parameter $\sigma$. The invariant charges $\Z^\pm_{\mu_1\dots
\mu_N}$ are unaffected by translations and transform covariantly
under Lorentz transformations in the ambient Minkowski space. They
form a Poisson algebra. This algebra can be identified with the
tensor product of two Poisson algebras -- one formed by the invariant
charges $\Z^+_{\mu_1\dots\mu_N}$, the other forned by the invariant
charges $\Z^-_{\mu_1\dots\mu_N}$ -- which ``Poisson commute'' with
each other. As
multiplicative algebras, the two subalgebras are isomorphic and are
generated by so-called standard invariants \cite{AlgProp}: the first
subalgebra is generated by standard invariants $\Z^{+(K)}_{\mu_1\dots
\mu_N}$, the second by standard invariants $\Z^{-(K)}_{\mu_1\dots
\mu_N}$ for certain values of $K$: $1\le K \le \frac{N+1}2$, and of
$\mu_1\dots\mu_N$. As Lie
algebras, with the Poisson bracket operation as the composition law,
they differ by a global factor $-1$ for the structure constants. Thus
it suffices to analyze just one of the two subalgebras.
     
For definiteness, choose the Poisson algebra of the invariant (internal) 
charges $\Z^{+(K)}_{\mu_1\dots\mu_N}$. Take advantage of the fact that
the generators of translations
\[  \P_\mu = \oint\limits_{\tau\,{\rm fixed}} \!\! \d\sigma \:
             u_\mu^+ (\tau,\sigma)     \]
are central elements of this algebra by treating them as c--numbers
$p_\mu$. Specialize to vectors $p_\mu$ on an arbitrary, yet fixed
mass shell with mass $\gm > 0$ and positive energy.

Exploit Lorentz invariance of the full \aoo\ and pass to the rest
frame of $p_\mu$:
\[  \P_\mu = \gm \, \delta_\mu^0  .\]
Denote the \a\ obtained in this way by $\gh$. $\gh$ is an \a\ with a
$*$--operation given by complex conjugation.

Within this \a\ $\gh$ the infinitesimal generators of rotations (in
the momentum rest frame) are given w.r.t.\ a Lorentzian basis by
\[  \JKT_i = \sum_{j=1}^3 \sum_{k=1}^3 \frac{-2\pi\ap}{4\gm} \varepsilon
            _{ijk} \Z^{+(2)}_{0jk} ,\qquad\qquad  i=1,2,3 . \]
After multiplication by the imaginary unit, they are to be inserted
into the first argument of the Poisson bracket $\{ \,,\, \}^{\rm PB
}$. The $\JKT_i$'s are the left mover parts of the components of the
spin vector.

Using the angular momentum basis instead of the Lorentzian basis, the
triplet
\[  \JT_{1,-1} := \frac1{\sqrt{2}} \, \JKT_- , \qquad
    \JT_{1,0}  := \JKT_3 ,\qquad
    \JT_{1,+1} := \frac{-1}{\sqrt{2}} \, \JKT_+   \]
with
\[  \JKT_\pm := \left( \JKT_1 \pm i \JKT_2 \right) \]
forms itself an irreducible tensor variable
\[  \JT_1 = \left\{ \JT_{1,m} \Big| m=-1,0,+1 \right\} . \]
Here, in analogy to the irreducible tensor operator, an irreducible
tensor variable
\[  \O_j = \Big\{ \O_{j,m} \Big| m = -j,-j+1,\dots,+j-1,+j \Big\} \]
-- built from the left movers and carrying (integer) spin
index $j\ge 0$ -- is so defined that it satisfies the Poisson bracket
relations
\[  \Big\{ i \JKT_3 , \O_{j,m} \Big\}^{\rm PB} = m \, \O_{j,m} ,\qquad
    \Big\{ i \JKT_\pm , \O_{j,m} \Big\}^{\rm PB} =
    \sqrt{(j\pm m+1)(j\mp m)} \; \O_{j,m\pm 1} .\]
$\JT_1$ carries the dimension of an action. The values of its components
are appropriately stated in terms of complex multiples of a unit given
by a typical action ${\bf A}$.

I replace the Poisson bracket $\{ \,,\,\}^{\rm PB}$, behaving as far as
dimensions are concerned like an inverse action, by the rescaled Poisson
bracket
\[  \Big\{ \,,\, \Big\}_{\rm PB} = 2\pi\ap \Big\{ \,,\,\Big\}^{\rm PB}, \]
behaving dimensionally like a mass to the power minus two. As long as
there is no risk of confusion with the anticommutator or the set
theoretic symbol, I shall drop the subscript PB and simply denote
the rescaled Poisson bracket by the symbol $\{\,,\,\}$.

With respect to the rescaled Poisson bracket $\{\,,\,\}$ the infinitesimal
generators of rotations in the momentum rest frame are given by
\[  \JK_i = \frac{-1}{4\gm} \varepsilon_{ijk} \Z^{+(2)}_{0jk}
    \quad \Big( := \sum_{j=1}^3 \sum_{k=1}^3 \frac{-1}{4\gm}
    \varepsilon_{ijk} \Z^{+(2)}_{0jk} \Big),   \qquad\qquad\qquad
    i=1,2,3 . \]
With the corresponding irreducible tensor variable
\[  \J_1 = \Big( \J_{1,-1} , \J_{1,0} , \J_{1,+1}  \Big) \]
the Poisson bracket relations are
\be  0 = i \Big\{ \J_{1,-1},\J_{1,+1} \Big\} - \J_{1,0} ,\qquad\qquad\qquad
     0 = i \Big\{ \J_{1,0} ,\J_{1,\pm 1} \Big\} \mp \J_{1,\pm 1} ;
     \label{1} \ee
\be  0 = i \Big\{ \J_{1,0}, \O_{j,m} \Big\} - m \O_{j,m}  , \quad\;
     0 = i \Big\{ \J_{1,\pm 1}, \O_{j,m} \Big\} \pm \frac{1}{\sqrt{2}}
         \, \sqrt{(j\pm m+1)(j\mp m)} \:\; \O_{j,m\pm1}.
     \label{2} \ee
The values of the dynamical quantities $\J_{1,m}$ are appropriately 
stated as complex multiples of the unit
\[  \left( \frac{{\bf A}}{2\pi\ap} \right) = \left(
    \frac{{\bf A}}{2\pi\ap \gm^2} \right) \times \gm^2 , \]
$( {\bf A} / 2\pi\ap\gm^2 )$ being dimensionless.
Actually, for any given invariant charge $\Z^{+(K)}_{
\mu_1\dots\mu_N}$ the appropriate scale is
\[  \left( \frac{{\bf A}}{2\pi\ap} \right)^{N-K} \times \gm^{2K-N} =
    \left( \frac{{\bf A}}{2\pi\ap\gm^2} \right)^{N-K} \times \gm^N . \]
The power of the typical action ${\bf A}$ enters into the ``degree''
\[   l := N-K-1  \]
assigned to the invariant charge $\Z^{+(K)}_{\mu_1\dots\mu_N}$. Poisson
bracket operation $\{ \Z^{+(K_1)}_{\lambda_1\dots\lambda_{N_1}}, \Z^{+
(K_2)}_{\mu_1\dots\mu_{N_2}} \}$ yields a linear combination of invariant
charges $\Z^{+(K_1+K_2-1)}_{\nu_1\dots\nu_{N_1+N_2-2}}$ with coefficients
from the real and complex numbers for a Lorentzian basis and an angular
momentum basis, respectively. The degree $l$ behaves additively under
Poisson bracket operation:
\[  \mbox{degree} \Big( \Big\{ \Z^{+(K_1)}_{\lambda_1\dots\lambda_{N_1}},
      \Z^{+(K_2)}_{\mu_1\dots\mu_{N_2}} \Big\} \Big) =
    \mbox{degree} \Big( \Z^{+(K_1)}_{\lambda_1\dots\lambda_{N_1}} \Big) +
    \mbox{degree} \Big( \Z^{+(K_2)}_{\mu_1\dots\mu_{N_2}} \Big) ,\]
in contrast to its behaviour under ordinary multiplication:
\[  \mbox{degree} \Big( \Z^{+(K_1)}_{\lambda_1\dots\lambda_{N_1}} \cdot
      \Z^{+(K_2)}_{\mu_1\dots\mu_{N_2}} \Big) =
    \mbox{degree} \Big( \Z^{+(K_1)}_{\lambda_1\dots\lambda_{N_1}} \Big) +
    \mbox{degree} \Big( \Z^{+(K_2)}_{\mu_1\dots\mu_{N_2}} \Big) + 1 . \]
As it will turn out in the following section, this distinct behaviour
is vital for the quantization, \ie\ the passage from the Poisson \a\
$\gh$ to the associative (non-commutative) \a\ $\widehat{\gh}$.

Returning to the \a\ of classical \obs s $\gh$, this very same degree
$l$ endows $\gh$ with an $\N_0$--grading with respect to the Lie
structure:
\[  \gh = \bigoplus\limits_{l=0}^\infty \V^l . \]
The homogeneous subspaces $\V^l$, $l=0,1,2,\dots$, are invariant
under the $*$--operation. They are finite dimensional. For instance,
the vector space $\V^0$ is three-dimensional, the vector spaces
$\V^1$, $\V^2$ and $\V^3$ are 20-, 92- and 468-dimensional,
respectively.

With respect to the second underlying structure of a Poisson \a,
the multiplication, $\gh$ is freely ({\it not} finitely) generated
by the standard invariants:
the number $n_0$ of generators contained in $\V^0$ being three,
the number $n_l$ of generators contained in each $\V^l$, $l\ge 1$,
being given by the formula
\[  n_l = n(4,l+2) - n(4,l+1)  \]
where
\[  n(4,N) := \frac1{N} \sum_{D|N} \mu(D) \: 4^{N/D} .\]
Here the sum extends over all divisors $D$ of $N$. The symbol $\mu
(D)$ denotes the M\"obius function of $D$:
\[  \mu(D) := \left\{  \begin{array}{cl}
    1 & \mbox{if $D=1$}, \\
    (-1)^p & \mbox{if $D$ can be decomposed into exactly $p$ different
                   prime factors,} \\
    0 & \mbox{if some prime factors of $D$ coincide.}
    \end{array} \right. \]
Consequently, the dimension of $\V^l$ for $l\ge 1$ is given by a sum
(over the number $q$ of generators occurring in the linearly independent
monomials contained in $\V^l$) of terms $t^{(q)}_l$
\[  \mbox{dim} \,\left( \V^l \right) = \sum_{q=1}^{l+1} t_l^{(q)},
    \qquad\qquad\qquad\quad   l \ge 1 \]
with
\[  t^{(q)}_l =
    \underbrace{\sum_{q_0=0}^l \sum_{q_1=0}^l \cdots \sum_{q_l=0}^l}_
    {q_0 +\dots+ q_l = q \atop 1\cdot q_1 + 2\cdot q_2 +\dots+ l\cdot
     q_l = l+1-q }
    { n_0 + q_0 - 1 \choose q_0 } { n_1 + q_1 - 1 \choose q_1 }
    \cdots { n_l + q_l - 1 \choose q_l }.  \]
In this sense, the dimension of the vector spaces $\V^l$ is under 
control.

The vector spaces $\V^l$ can be decomposed into a direct sum of 
their positive parity and their negative parity parts:
\[  \V^l = \V^l_+ \oplus \V^l_-  .\]
Actually, $\V^0_-$ is the set $\{0\}$. $\V^0_+$ considered as a 
vector space coincides with the linear span of the invariant
charges $\JK_i$, $i=1,2,3$, or, equivalently, of the invariant
charges $\J_{1,m}$, $m=-1,0,+1$. 

Considered as an algebra w.r.t.\ Poisson bracket operation
(Lie--Poisson \a), $\V^0_+$ coincides with the Lie \a\
$so(3)$ of the infinitesimal generators of rotations in the
momentum rest frame
\[  \Big\{ i \JK_i,\JK_j \Big\} = i \,\varepsilon_{ijk} \JK_k .\]
Moreover, every vector space $\V^l_+$ and $\V^l_-$ is a \r\ space
for the Lie--Poisson \a\ $\V^0_+$. Consequently, $\V^l_+$ and
$\V^l_-$ can be decomposed into a direct sum of isotypical components
corresponding to the spin $j$: $\V^l_{j,+}$ and $\V^l_{j,-}$, respectively.
The index $j$ takes integer values between 0 and $l+1$.

Using this notation, for example $\V^0_+$, $\V^1_+$ and $\V^1_-$
can be decomposed as
\[  \V^0_+ = \V^0_{1,+} ,\qquad 
    \V^1_+ = \V^1_{0,+} \oplus \V^1_{2,+} ,\qquad
    \V^1_- = \V^1_{1,-} \oplus \V^1_{2,-}  \]
with pertinent dimensions 3, $2\oplus10$ and $3\oplus5$, respectively.

The subspace $\V^1_{0,+}$ is spanned by the elements 
\[  \big( \J^2_1 \big)_0  :=  \langle 0,0 | 1,m_1;1,m_2 \rangle
    \; \J_{1,m_1} \J_{1,m_2} \quad
    \Big( := \sum_{m_1}\sum_{m_2} \, \langle 0,0 | 1,m_1;1,m_2 \rangle
    \; \J_{1,m_1} \J_{1,m_2} \Big)    \]
and
\[  \B_0^{(1)}  :=  \frac12 \sum_{j=1}^3  \Z^{+(2)}_{0j0j} ,\]
the subspace $\V^1_{2,+}$ by the components of the \ir\ tensor
variables
\[  \left( \J_1^2 \right)_2  =  \Big\{ \left( \J_1^2 \right)_
    {2,m}  \Big|  m=-2,\dots,+2  \Big\}  \]
and
\[  \T_2  =  \Big\{ \T_{2,m} \Big| m=-2,\dots,+2  \Big\}  \]
with
\[  \left( \J_1^2 \right)_{2,m}  :=  \langle 2,m | 1,m_1;1,m_2 
    \rangle  \J_{1,m_1} \J_{1,m_2},  \]
\[  \T_{2,-2}  :=  \frac{-1}8  \Z^{+(2)}_{0101} + \frac18
    \Z^{+(2)}_{0202} + \frac14 \Z^{+(2)}_{0102}  \]
and $\T_{2,m}$ obtained by repeated action of $i \JK_+$ on $\T_{2,-2}$
according to formulae (\ref{1}) and (\ref{2}), the subspace $\V^1_{1,-}$ 
is spanned by the components of the \ir\ tensor variable
\[  \S_1 = \Big\{ \S_{1,m} \Big| m=-1,0,+1 \Big\}  \]
with
\[  \S_{1,-1}  :=  \frac1{2\sqrt{2}} \sum_{j=1}^3 \left(
    \Z^{+(2)}_{0j1j} - i \Z^{+(2)}_{0j2j}  \right)  \]
and $\S_{1,m}$ obtained by repeated action of $i\JK_+$ on $\S_{1,-1}$
according to formulae (\ref{1}) and (\ref{2}), and, finally, the subspace 
$\V^1_{2,-}$ by the components of the \ir\ tensor variable
\[  \S_2  =  \Big\{ \S_{2,m} \Big| m=-2,\dots,+2 \Big\}  \]
with
\[  \S_{2,-2}  :=  \frac{-i}4 \Z^{+(2)}_{0131} + \frac{i}4 
    \Z^{+(2)}_{0232} - \frac12 \Z^{+(2)}_{0132}  \]
and $\S_{2,m}$ obtained by repeated action of $i\JK_+$ on $\S_{2,-2}$
according to formulae (\ref{1}) and (\ref{2}). 

The symbols $\langle j,m 
| j_1,m_1;j_2,m_2 \rangle$ above denote the Clebsch--Gordan coefficients 
of Condon and Shortley. The normalization of the \ir\ tensor variables
$\O_{j,m}$ is such that
\be  0 = \O_{j,m}^* - (-1)^m \O_{j,-m}  .\label{3} \ee
Nota bene: the tensor variables $\J_1$, $\B_0^{(1)}$ and $\T_2$
carry positive parity, whereas the tensor variables $\S_1$ and
$\S_2$ carry negative parity.

Alternatively, the vector spaces $\V^l_+$ and $\V^l_-$ can be decomposed
into direct sums of eigenspaces of $\{ i\JK_3, \cdot \}_{\rm PB}$
corresponding to eigenvalues $m \in \{ -(l+1),\dots,+(l+1) \}$
\[  \V^l_+ = \bigoplus_m \V^l_{+,m} ,\qquad\qquad
    \V^l_- = \bigoplus_m \V^l_{-,m} .\]
This allows to compute successively the dimension of every subspace
$\V^l_{j,+}$, $0\le j\le l+1$, as $(2j+1)$ times the difference between
the dimension of $\V^l_{+,j}$ and the dimension of $\V^l_{+,j+1}$,
beginning with $j=l+1$ and dim$(\V^l_{+,l+2})=0$. The same goes for 
the computation of the dimensions of the subspaces $\V^l_{j,-}$, $0
\le j\le l+1$. Note that the dimensions of $\V^l_{+,0}$ and $\V^l_{
-,0}$ coincide with the number of irreducible spin multiplets in $\V^l_+$
and $\V^l_-$,
respectively. By turns, in essentially the same manner as the dimension
of $\V^l$ was computed before, the dimensions of $\V^l_{+,m}$ and
$\V^l_{-,m}$, $-(l+1)\le m\le +(l+1)$, can be computed from the numbers
$n_{l^\prime,+,m^\prime}$ and $n_{l^\prime,-,m^\prime}$ of the generators
of the symmetric \a\ (\ie\ the standard invariants)
which are contained in the subspaces $\V^{l^\p}_
{+,m^\p}$ and $\V^{l^\p}_{-,m^\p}$, $l^\p \le l$, $-(l^\p +1) \le m^\p
\le +(l^\p +1)$.

In their turn, these numbers can be obtained from the following
formulae
\[  n_{l,+,m}  =  \underbrace{ \sum_{n_0 = l,l-2,l-4,\dots \ge 0}
                  \sum_{n_1=0}^{l+1} \sum_{n_2=0}^{l+1} \sum_{n_3=0}^{l+1} 
                  }_{ n_0+n_1+n_2+n_3=l+2 \atop  n_1-n_2=m }
                  M_0^\p (n_0,n_1,n_2,n_3) , \qquad
                  {-l-1 \le m \le l+1 \atop l=0,1,2,3,\dots}  \]
\[  n_{l,-,m}  =  \underbrace{ \sum_{n_0 = l-1,l-3,\dots \ge 0}
                  \sum_{n_1=0}^{l+1} \sum_{n_2=0}^{l+1} \sum_{n_3=0}^{l+1} 
                  }_{ n_0+n_1+n_2+n_3=l+2 \atop  n_1-n_2=m }
                  M_0^\p (n_0,n_1,n_2,n_3) , \qquad
                  {-l-1 \le m \le l+1 \atop l=1,2,3,\dots}  \]
with
\[  M_0^\p(n_0,n_1,n_2,n_3)  =  \left\{ \begin{array}{ll}
    M_0(n_0,n_1,n_2,n_3) - M_0(n_0-1,n_1,n_2,n_3) & \;\mbox{for $n_0\ge1$},\\
    M_0(0,n_1,n_2,n_3) & \;\mbox{for $n_0=0$.}  \end{array} \right.  \]
Here the symbol $M_0(k_1,k_2,\dots,k_r)$, $k_1,\dots,k_r \in \N$,
denotes the following function
\[  M_0(k_1,k_2,\dots,k_r)  =  \frac1{K} \sum_{D|k_\nu ,\, \nu=1,2,\dots,r}
    \frac{\mu(D) (K/D)!}{\prod_{\nu=1}^r (k_\nu /D)!}  ,\qquad\qquad
    K  :=  \sum_{\nu=1}^r k_\nu .  \]
\vspace{3mm}

To begin with the detailed structural analysis, I consider $\gh$
merely as an \a\ with respect to multiplication, a freely generated
one. I divide its generators (organized in irreducible tensor variables,
in the sequel called irreducible tensor generators)
into two disjoint sets:

\begin{itemize}

\item[{\it i})]
the set of real, algebraically independent, positive parity, scalar
generators $\B_0^{(l)}$, $l=1,3,5,\dots$, of degree $l$, the linear
span of which forms an abelian Lie \a\ $\ga$, while the polynomials
in $\B_0^{(l)}$ form a maximal abelian Lie sub\a\ of the (Poisson)
\a\ $\gh$ \cite{?}, and

\item[{\it ii})]
a complementary set of suitably chosen \ir\ tensor generators, each one
of them
carrying a well-defined degree $l\ge0$. The polynomials of all these
generators form a sub\a\ $\U$ of $\gh$ with respect to multiplication.

\end{itemize}
Clearly, $\ga \cap \U = \{0\}$ and, clearly, $\gh$ coincides with the
polynomial ring in the elements of $\ga$ and $\U$, the product in the
algebra $\U$ and the product in the polynomial ring being identified.

At this point, the Lie sub\a\ $\ga$ is not uniquely defined, yet, in
contrast to the maximal abelian Lie sub\a\ of the (Poisson) \a\ $\gh$,
generated by it through multiplication. Also the subalgebra $\U$ is not
uniquely defined at this point. This non-uniqueness will be
resolved shortly. In any case the subalgebra $\ga$ as a whole carries
positive parity while $\U$ contains elements of either parity:
\beas  \ga &=& \ga_+ , \\
       \U  &=& \U_+ \oplus \U_- . \eeas
The ``grade'' $p=l+1$ endows $\ga=\ga_+$ and $\U$ with an $\N$--gradation
with respect to multiplication:
\[  \ga = \bigoplus_{p=2,4,\dots} \ga_+^{(p)} , \qquad\qquad
    \U  = \bigoplus_{p=1,2,\dots} \U^{(p)}
        = \bigoplus_{p=1,2,\dots} \left( \U_+^{(p)} 
          \oplus \U_-^{(p)} \right) .  \]
By means of this gradation, I define
\beas  \ga_+^l   &=&  \ga_+^{(l+1)}  \qquad\qquad\: 
                      \mbox{for $l=1,3,5,\dots$} \,,  \\
       \U^l_\pm  &=&  \U^{(l+1)}_\pm  \qquad\qquad
                      \mbox{for $l=0,1,2,\dots$ \,.}    \eeas
Now, employing solely that part of the Lie structure of $\gh$ which
is necessary to define the action of the generators $\JK_k$, $k=1,2,
3$, on the elements of $\gh$ via $\{ i\JK_k,\cdot \}_{\rm PB}$, I
observe that each one of the subspaces $\ga^l_+$, $\U^l_+$ and $\U^l
_-$ is a \r\ space for $so(3) \cong \U_+^0 = \V^0$ ($\U^0_- = \{0\}$
). Hence, each one of these subspaces can be decomposed into invariant
subspaces corresponding to equivalent \ir\ \r s of $so(3)$ labelled by
the spin-value $j$ with $0\le j \le l+1$, or, alternatively, into 
eigenspaces of $i\JK_3$ corresponding to the eigenvalue $m$ with $-(l+1)
\le m \le +(l+1)$. For the one-dimensional subspaces $\ga^l_+$ both
decompositions are trivial, while for $\U^l_+$ and $\U^l_-$ they are
highly non-trivial and helpful. As before, the $m$--decomposition
provides the information about the dimension of the $j$--decomposition.
Here are some examples:
\[  \!\!\!\!\!\!\!\!\!\!\!\!\!\!\!\!\!\!\!\!\!\!\!\!\!\!\!
 l=1: \qquad  \U^1_+ = \U^1_{0,+} \oplus \U^1_{2,+}  \qquad
         \mbox{and} \qquad \U^1_- = \U^1_{1,-} \oplus \U^1_{2,-}  \]
with respective dimensions $1\oplus 10$ and $3\oplus 5$;
\[  \;\;\,
    l=2: \qquad  \U^2_+ = \U^2_{1,+} \oplus \U^2_{2,+} \oplus \U^2_{3,+}
         \qquad \mbox{and} \qquad 
         \U^2_- = \U^2_{0,-} \oplus\dots\oplus \U^2_{3,-}  \]
with respective dimensions $15\oplus 10\oplus 21$ and $2\oplus 12
\oplus 15\oplus 14$;
\[  l=3: \qquad  \U^3_+ = \U^3_{0,+} \oplus\dots\oplus \U^3_{4,+}
         \qquad \mbox{and} \qquad
         \U^3_- = \U^3_{0,-} \oplus\dots\oplus \U^3_{4,-}  \]
with respective dimensions $11\oplus 30\oplus 90\oplus 42\oplus 54$
and $6\oplus 45\oplus 70\oplus 63\oplus 36$.
Actually, $\U^1_{2,+} \equiv \V^1_{2,+}$, $\U^1_{1,-} \equiv \V^1_{
1,-}$, $\U^1_{2,-} \equiv \V^1_{2,-}$, while $\U^1_{0,+}$ is the
complex line $\{ \lambda \cdot (\J^2_1)_0  |  \lambda \in \C \}$.

Equally important as the knowledge of the dimensions of the various
vector spaces $\U^l_{j,+}$ and $\U^l_{j,-}$, is the knowledge of the
numbers $\bar{n}_{l,j,+}$ and $\bar{n}_{l,j,-}$, $0\le j \le l$, $l
=1,2,\dots$, of the generators contained in $\U^l_{j,+}$ and $\U^l_
{j,-}$, respectively, which generate the ``multiplicative'' \a\ $\U$
freely. This information is provided by the formulae ($l=1,2,3,\dots$)
\[  \bar{n}_{l,j,-} := (2j+1)(n_{l,-,j} - n_{l,-,j+1})  \qquad\quad
    \mbox{for $j=0,1,\dots,l+1$}  \]
with $n_{l,-,l+2} = 0$, and
\[  \bar{n}_{l,j,+} := (2j+1) \times \left\{ \begin{array}{lcl}
    (n_{l,+,j} - n_{l,+,j+1}) & \qquad & \mbox{for $j=1,\dots,l+1$}, \\
    (n_{l,+,0} - n_{l,+,1}) & & \mbox{for $j=0$, $l=0$ mod 2}, \\
    (n_{l,+,0} - n_{l,+,1} - 1) & & \mbox{for $j=0$, $l=1$ mod 2}
    \end{array} \right.  \]
with $n_{l,+,l+2} = 0$.

To proceed further I shall {\it assume} that the following conjecture
holds true:

\begin{itemize}

\item[{\it i})]
$\U$ is a {\it Poisson} subalgebra of $\gh$. (In this case the
$l$--decomposition
$\U = \bigoplus_{l=0}^\infty \U^l = \bigoplus_{l=0}^\infty (\U^l_+
\oplus \U^l_-)$ corresponds to an $\N_0$--gradation of $\U$ with respect
to its Lie structure.)

\item[{\it ii})]
The derived \a
\[  \Big\{ \bigoplus_{l=1}^\infty \U^l , \bigoplus_{l=1}^\infty \U^l
    \Big\} \subset \bigoplus_{l=2}^\infty \U^l  \]
contains all generators of $\U$ with degree $l>1$, $\U$ being considered
here solely as an algebra with respect to multiplication.

\item[{\it iii})]
The generators $\B^{(l)}_0$, $l=1,3,5,\dots$, which span the abelian 
subalgebra $\ga\subset\gh$, can be chosen such that the sum $\gg$ of
$\ga$ and $\U$ as {\it Lie} algebras is semi-direct, and such that
they act as derivations on the {\it Poisson} algebra $\U$
\[  \gg = \ga \ltimes \U  . \]
\end{itemize}
This conjecture has been verified for $l\le 7$, partly with the help
of a computer program for symbolic computations.

Item $ii)$ of the conjecture implies that for $l\ge2$ the $\bar{n}_{
l,j,+}$ generators and the $\bar{n}_{l,j,-}$ generators of the 
multiplicative 
algebra $\U$, which are contained in the subspaces $\U^l_{j,+}$ and
$\U^l_{j,-}$, respectively, are obtainable in the form of algebraically
independent iterated Poisson brackets with exactly $l$ entries from
the vector spaces $\U^1_{j,+}$ and $\U^1_{j,-}$.

This is tantamount to saying that, as a Poisson algebra, $\U$ is
finitely generated by the \ir\ tensors $\J_1 \in \U^0_{1,+}$, $\T_2
\in \U^1_{2,+}$, $\S_1 \in \U^1_{1,-}$ and $\S_2 \in \U^1_{2,-}$,
or rather from two components of $\J_1$ and one component each of
$\T_2$, $\S_1$ and $\S_2$. This resolves the previous non-uniqueness
in the choice of $\U$.

Item $iii)$ of the conjecture resolves the previous non-uniqueness
in the choice of $\ga$. $\gh$ is the symmetric algebra over $\gg$,
the product in $\U$ and the product in the symmetric algebra
being identified.

For a presentation of the Poisson algebra $\gh$ it suffices to present
the Poisson algebra $\U$ and to specify the action of the basis
elements of the abelian algebra $\ga$ on the generators of $\U$.

$\U$ will be presented as a two-fold quotient of the free 
{\it Poisson} algebra $\F_0$ with abstract generators $\SJ_{1,m}$,
$\ST_{2,m}$, $\SS_{1,m}$ and $\SS_{2,m}$ carrying the same dimensions 
and degrees as in the Nambu--Goto theory. Besides, $\F_0$ is equipped
with a $*$--operation and a parity transformation $P$ which operate on
the generators of $\F_0$ as follows:
\[  0 = (\SJ_{1,m})^* - (-1)^m \,\SJ_{1,-m} ,\qquad
    0 = (\ST_{2,m})^* - (-1)^m \,\ST_{2,-m} ,\]
\[  0 = (\SS_{1,m})^* - (-1)^m \,\SS_{1,-m} ,\qquad
    0 = (\SS_{2,m})^* - (-1)^m \,\SS_{2,-m} ,\]
\[  0 = (\SJ_{1,m})^P - \SJ_{1,m} ,\quad
    0 = (\ST_{2,m})^P - \ST_{2,m} ,\quad
    0 = (\SS_{1,m})^P + \SS_{1,m} ,\quad
    0 = (\SS_{2,m})^P + \SS_{2,m}  \]
and are extended to all of $\F_0$ as a $*$--operation and an
automorphism, respectively. Moreover $\F_0$ is endowed with
two gradings

\begin{itemize}

\item[{\it a})]
with respect to the multiplicative structure given by the ``grade''
($l+1$);

\item[{\it b})]
with respect to the Lie structure given by the ``degree'' $l$.

\end{itemize}

The first quotient of $\F_0$ is performed with respect to the graded
ideal $\I_0$. This quotient takes care of

\begin{itemize}

\item[{\it i})]
the Poisson bracket relations of the components of $\J_1$ among each 
other,

\item[{\it ii})]
the Poisson bracket relations of the components of $\J_1$ with the
components of $\T_2$, $\S_1$ and $\S_2$.

\end{itemize}
In other words, $\I_0$ is generated by the following linear combinations:
the right hand sides of the equations (\ref{1}) and (\ref{2})
(with the generators $\SJ_1$ and $\ST_2$, $\SS_1$, $\SS_2$
substituted for the generators $\J_1$ and $\O_{j,m}$,
respectively).
All these generating elements are homogeneous in the degree $l$, 
in the spin value $j$, in the sign of the parity and in their reality 
properties. Hence the ideal $\I_0$ is endowed with the same two-fold
grading as $\F_0$ and it is invariant under rotations in the momentum
rest frame, under conjugation and parity transformation.

The result of the above identifications is the Poisson--$*$--algebra
\[  \F = \F_0 \Big/ \I_0   \]
with generators $J_1$, $T_2$, $S_1$ and $S_2$. $\F$ is equipped with 
a two-fold grading and it is invariant under rotations in the momentum
rest frame, under reflections and under conjugation, the latter operations
being well-defined in $\F$. The generators
$J_1$, $T_2$, $S_1$ and $S_2$ carry the same dimensions and degrees
as the respective variables in the Nambu--Goto theory.

The second and much more specific quotient of $\F_0$ is taken with 
respect to an ideal $\I$ generated by some particular 
polynomials in the generators $J_1$, $T_2$, $S_1$ and $S_2$ of $\F$
and in their iterated Poisson brackets. This quotient takes care of
the polynomial relations valid in the classical Nambu--Goto theory
in the corresponding variables, polynomial relations other than
those expressing the derivation property, antisymmetry property 
and Jacobi identity of the Poisson bracket and other than those
expressing the covariance of the Poisson bracket and of the 
generators under rotations in the momentum rest frame, under 
reflections and under conjugation. This quotient must account
for the difference of the dimensions of the (degree $l$, spin
value $j$, parity plus or minus) subspaces $\F^l_{j,\pm}$ of
$\F$ and the corresponding subspaces $\U^l_{j,\pm}$ of $\U$
in the classical Nambu--Goto theory:

The dimensions of the subspaces $\F^l_{j,\pm}$ are computed 
essentially in the same manner as those of $\U^l_{j,\pm}$, this
time with the numbers $N^l_{\pm,m}$ of generators (w.r.t.\ 
multiplication) contained in each subspace $\F^l_{\pm,m}$,
$l\ge 1$ (corresponding to the $m$--decomposition of $\F^l
_\pm$) obtained from the Witt formula applied to thirteen
$l=1$ generators, {\it viz}.\ $T_{2,m}$, $S_{1,m}$, $S_{2,m}$:
\[  N^l_{{\scriptscriptstyle + \atop (-)},m} = \underbrace{ \sum_{n_1 \ge 0} 
    \sum_{n_2 \ge 0} \cdots\cdots\cdots \sum_{n_{13} \ge 0}}_
    { n_1+n_2+\dots\dots\dots\dots\dots+n_{12}+n_{13}=l \atop 
    { n_1+\dots+n_5-l=0 \, mod \; 2 \;\; (=1 \, mod \; 2) \atop 
      -2(n_1+n_9)-n_2-n_6-n_{10}+n_4+n_8+n_{12}+2(n_5+n_{13})=m}} 
    \!\!\!\!\!\!\!\!\!\! M_0(n_1,\dots,n_{13}). \]
The concrete number and shape of the generators of the ideal $\I$
is obtained from a basis of additional independent homogeneous
polynomial relations in the variables $\J_{1,m}$, $\T_{2,m}$, $\S
_{1,m}$, $\S_{2,m}$ and their iterated Poisson brackets, valid in
the classical Nambu--Goto theory, without exploitation of the 
existence of the scalar elements $\B_0^{(l)}$, $l=1,3,5,\dots$,
their commutativity,
and their specific actions on the generators $\T_{2,m}$, $\S_{1,m}$
and $\S_{2,m}$. Without loss of generality, these ``$\U$--generating
relations''can be arranged to be homogeneous in the degree $l$, in
the spin value $j$, in the sign of the parity and in the behaviour 
under conjugation, and can be grouped into \ir\ multiplets.

\bs
If, with the help of certain multiplets of polynomials $\Q^l_{j,\pm,r}
(z_1,\dots,z_s)$, $r=1,2,3,\dots$, in the indeterminates $z_1,\dots,
z_s$, $s=1,2,3,\dots$, these $\U$--generating relations of the
classical Nambu--Goto theory are cast into the form
\[  0 = \Q^l_{j,\pm,r}(z_1,\dots,z_s)  \]
for a definite identification of the indeterminates $z_i$ with the
variables $\J_{1,m},\dots,\S_{2,m}$ and their (iterated) Poisson
brackets, then the generators of $\I$ are given by $\Q^l_{j,\pm,r}
(z_1,\dots,z_s)$ with the corresponding identification of the 
indeterminates $z_i$ with the generators $J_{1,m},\dots,S_{2,m}$
of $\F$ and their (iterated) Poisson brackets.
\es

The pertinent computations of the $\U$--generating relations 
of the Nambu--Goto theory are performed with the help of the 
``modified Poisson brackets'' for the building blocks $\R^t_
{\mu_1\dots\mu_N}$ of the invariant charges $\Z^{+(K)}_{\mu_1
\dots\mu_N}$ \cite{AlgProp}. For invariant charges (expressed
as polynomials in the building blocks $\R^t_{\mu_1\dots\mu_N}$)
the original Poisson bracket operation and the modified Poisson
bracket operation give identical results. However, the use of
the modified Poisson brackets offers considerable practical advantages
over the use of the original Poisson brackets.

Obviously, there are no such relations for $l=0$ and $l=1$.
The complete list of $\U$--generating relations for $l=2$, 3
and 4 is given below. For $l=4$ there are only 4 such relations
and even these ones are not independent of the $l=2$ and $l=3$
relations. In fact, they are all induced by the Poisson bracket
operation of the scalar element $\B_0^{(1)}$ from the $l=2$ and
$l=3$ relations. Moreover, the emergence of ``{\it truly independent}''
generating relations for $l>5$ is rather unlikely, particularly
in view of the growing assistance of more and more elements $
\B_0^{(l)} \in \ga$ in the induction procedure via their Poisson
bracket operation on the $\U$--relations with lower degree and
in view of the growing number of induced $\U$--relations
in the wake of the abelian commutation relations of more and more
elements of $\B_0^{(l)} \in \ga$. In any case, the quotient $\U=
\F \big/ \I$ still contains a multitude of non-trivial ideals
w.r.t.\ either one of its underlying algebraic structures. Examples of 
such ideals are the set of all elements of $\U$ carrying degree $l\ge
k$ ($k=1,2,3,\dots$) and the polynomials of (ordinary) degree
$\ge k$ ($k=2,3,4,\dots$) in the generators $\J_1$, $\T_2$, $\S_1$,
$\S_2$ of $\U$ and (iterated) Poisson brackets thereof.

In order to settle the issue of completeness of the generators of 
$\I$ corresponding to a given set of $\U$--generating relations,
one might try to compare the dimensions of $\BI^l_{j,\pm}$, the
homogeneous subspaces of the ideal $\BI$ with only these generators,
$\BI\subset\I$, to the dimensions of $\F^l_{j,\pm}$ and $\U^l_{j,\pm}$.
For the computation of the dimensions of $\BI^l_{j,\pm}$, one could
exploit the existing explicit Poisson bracket isomorphism between
the graded ideal $\BI$ in the graded {\it Poisson algebra} $\F$ and 
a corresponding graded ideal $\bgi$ in the graded {\it Lie algebra} $\gf$
\[  \gf = \gf_0 \Big/ \gi_0  \]
where $\gf_0$ denotes the free {\it Lie} algebra with abstract
generators $\stackrel{\circ}{\SJ}_{1,m}$ ($l=0$), $\stackrel{\circ}
{\ST}_{2,m}$, $\stackrel{\circ}{\SS}_{1,m}$, $\stackrel{\circ}{\SS}_
{2,m}$ ($l=1$) and $\gi_0 \subset \gf_0$ the ideal defined in analogy
to the ideal $\I_0\subset\F_0$. The generators of the ideal $\bgi\subset
\gf$ are obtained from the generators of $\BI\subset\F$ by
deleting all non-trivial products contributing to the latter
(\ie\ leaving only the linear combinations of the iterated
Poisson brackets) and substituting the generators of $\gf$
for those of $\F$. The resulting generators for $\bgi$ are much
simpler than their counterparts for $\BI$. So one might hope 
that it is easier to keep track of the dependencies among the
elements of $\bgi$ than of the dependencies among the elements
of $\BI$. However, even in this simplified version, no practicable
algorithm is known at present which would lead to an analytic
counting formula for the relevant dimensions. Thus, comparison
of the dimensions of $\BI^l_{j,\pm}$, $\F^l_{j,\pm}$ and $\U^l_
{j,\pm}$ is not a viable option for settling the issue of
completeness of $\BI$. At present, the only -- albeit unsatisfactory
-- option seems to be a symbolic computer construction of the
homogeneous subspaces of the quotient algebra $\gf \big/ \bgi$
for values of $l$ ranging from 5 to 7. Computer programs of
this kind are available: see \eg\ Ref.\ \cite{Gerdt}. 

\vspace{2mm}

Here is the list of generating relations for the algebra
$\U$, formulated as non-identically satisfied relations in the
algebra $\F$. The notation is explained at the end of the list.
\begin{center}    \underline{$l=2$}   \end{center}
\[  \begin{array}{rcl}
\!\! J^P=4^-:\quad 0 &=& \{T_2,S_2\}_{4}\,; \vspace{1mm}\\
\!\! J^P=3^+:\quad 0 &=& \{T_2,T_2\}_{3} + 
           i\,\{S_2,S_1\}_{3} - i\,16\,(J_1^{3})_{3}\,; \vspace{1mm}\\
     0 &=& \{S_2,S_2\}_{3} - 
           i\,2\,\{S_2,S_1\}_{3} + 
           i\,8\,(J_1\cdot T_2)_{3} + i\,48\,(J_1^{3})_{3}\,; 
           \qquad\qquad\qquad\qquad\qquad
%          \vspace{1mm}\\
    \end{array}  \]
\[  \begin{array}{rcl}
\!\! J^P=3^-:\quad 0 &=& \{T_2,S_2\}_{3} - 
           i\,\{T_2,S_1\}_{3} - 
           i\,8\,(J_1\cdot S_2)_{3}\,; \vspace{1mm}\\
\!\! J^P=2^-:\quad 0 &=& \{T_2,S_2\}_{2} + 
           {\frac{i}{3}}\,{\sqrt{{\frac{7}{2}}}}\,
           \{T_2,S_1\}_{2} + 
           i\,{\frac{4}{3}}\,{\sqrt{14}}\,(J_1\cdot S_2)_{2}\,;
           \qquad\qquad\qquad\qquad\qquad\qquad \\
\!\! J^P=1^+:\quad 0 &=& \{S_2,S_2\}_{1} + 
           i\,{\sqrt{{\frac{2}{3}}}}\,\{S_2,S_1\}_{1} + 
           {\frac{1}{6}}\,{\sqrt{5}}\,\{S_1,S_1\}_{1} \\&&+ 
           i\,16\,{\sqrt{{\frac{2}{3}}}}\,(J_1\cdot T_2)_{1} + 
           i\,32\,{\sqrt{{\frac{2}{15}}}}\,
           (J_1\cdot (J_1^{2})_{0})_{1}\,;
   \end{array}  \]
\vspace{1mm}
\begin{center}    \underline{$l=3$}   \end{center}
\[   \begin{array}{ll} 
     \lefteqn{\!\! J^P=5^-:} \vspace{1mm}\\
\!\! 0 \;\; = & \{\{S_{2},S_{1}\}_{3},S_{2}\}_{5}\,;
     \qquad\qquad\qquad\qquad\qquad\qquad\qquad\qquad\qquad\qquad\qquad
     \qquad\qquad\qquad\qquad\qquad\qquad\qquad
     \end{array}  \]
\[   \begin{array}{ll}
     \lefteqn{\!\! J^P=4^+:}  \\
\!\! 0 \;\; = & \{\{S_{2},S_{1}\}_{2},T_{2}\}_{4} + 
     i\,{\frac{4}{9}}\,{\sqrt{{\frac{2}{3}}}}\,
     \{\{T_{2},S_{1}\}_{3},S_{1}\}_{4} + 
     i\,{\frac{40}{9}}\,{\sqrt{{\frac{2}{3}}}}\,
     (J_{1}\cdot \{S_{2},S_{1}\}_{3})_{4} \\
 &   + i\,{\frac{16}{3}}\,{\sqrt{{\frac{2}{3}}}}\,(T_{2}^{2})_{4} - 
     i\,4\,{\sqrt{{\frac{2}{3}}}}\,(S_{2}^{2})_{4} - 
     i\,{\frac{64}{3}}\,{\sqrt{{\frac{2}{3}}}}\,
     ((J_{1}^{2})_{2}\cdot T_{2})_{4} - 
     i\,{\frac{128}{3}}\,{\sqrt{{\frac{2}{3}}}}\,
     ((J_{1}^{2})_{2}^{2})_{4}\,;
     \qquad\qquad\qquad\qquad\qquad
     \end{array}  \]
\[  \begin{array}{ll}
    \lefteqn{\!\! J^P=4^-:}  \\
\!\!0 \;\; \stackrel{(*)}{=} & \{\{T_{2},S_{1}\}_{2},T_{2}\}_{4} + 
    {\frac{3}{2}}\,\{\{S_{2},S_{1}\}_{2},S_{2}\}_{4} - 
    i\,10\,{\sqrt{{\frac{2}{3}}}}\,(J_{1}\cdot \{T_{2},S_{1}\}_{3})_{4} 
    \qquad\qquad\qquad\qquad\qquad \\
 &  + i\,4\,{\sqrt{{\frac{2}{3}}}}\,(T_{2}\cdot S_{2})_{4}
    - i\,8\,{\sqrt{{\frac{2}{3}}}}\,((J_{1}^{2})_{2}\cdot S_{2})_{4}\,;
    \end{array}  \]
\[  \begin{array}{ll}
    \lefteqn{\!\! J^P=3^+:}  \\
\!\!0 \;\; \stackrel{(*)}{=} & \{\{S_{2},S_{1}\}_{2},T_{2}\}_{3} + 
    {\frac{1}{9}}\,{\sqrt{10}}\,\{\{S_{2},S_{1}\}_{1},T_{2}\}_{3} - 
    i\,{\frac{10}{27}}\,\{\{T_{2},S_{1}\}_{2},S_{1}\}_{3} \vspace{1mm}\\
 &  + i\,{\frac{260}{81}}\,{\sqrt{2}}\,(J_{1}\cdot \{S_{2},S_{1}\}_{3})_{3}
    - i\,{\frac{40}{81}}\,(J_{1}\cdot \{S_{2},S_{1}\}_{2})_{3} + 
    {\frac{212}{27}}\,{\sqrt{{\frac{2}{3}}}}\,(S_{2}\cdot S_{1})_{3}
    \qquad\qquad\qquad\qquad\qquad\\
 &  - i\,{\frac{32}{9}}\,{\sqrt{{\frac{2}{3}}}}\,((J_{1}^{2})_{2}
    \cdot T_{2})_{3}\,;\vspace{1mm}\\
\!\!0 \;\; \stackrel{(*)}{=} & \{\{T_{2},S_{1}\}_{3},S_{1}\}_{3} - 
    {\sqrt{2}}\,\{\{T_{2},S_{1}\}_{2},S_{1}\}_{3} + 
    {\frac{26}{3}}\,(J_{1}\cdot \{S_{2},S_{1}\}_{3})_{3} \vspace{1mm}\\
 &  + {\frac{16}{3}}\,{\sqrt{2}}\,(J_{1}\cdot \{S_{2},S_{1}\}_{2})_{3} + 
    i\,4\,{\sqrt{{\frac{1}{3}}}}\,(S_{2}\cdot S_{1})_{3} + 
    48\,{\sqrt{3}}\,((J_{1}^{2})_{2}\cdot T_{2})_{3}\,;
    \end{array}  \]
\[  \begin{array}{ll}   
    \lefteqn{\!\! J^P=3^-:}  \\
\!\!0 \;\; \stackrel{(*)}{=} & \{\{T_{2},S_{1}\}_{2},T_{2}\}_{3} + 
    {\frac{1}{3}}\,{\sqrt{10}}\,\{\{T_{2},S_{1}\}_{1},T_{2}\}_{3} - 
    {\frac{1}{6}}\,{\sqrt{{\frac{5}{2}}}}\,
    \{\{S_{2},S_{1}\}_{1},S_{2}\}_{3} \vspace{1mm}\\
 &  + i\,{\frac{10}{9}}\,{\sqrt{2}}\,\{\{S_{2},S_{1}\}_{3},S_{1}\}_{3} - 
    i\,{\frac{19}{36}}\,\{\{S_{2},S_{1}\}_{2},S_{1}\}_{3} + 
    i\,2\,{\sqrt{2}}\,(J_{1}\cdot \{T_{2},S_{1}\}_{3})_{3} 
    \qquad\qquad\qquad\qquad\qquad \vspace{1mm}\\
 &  - i\,{\frac{4}{9}}\,(J_{1}\cdot \{T_{2},S_{1}\}_{2})_{3} - 
    i\,28\,{\sqrt{{\frac{2}{3}}}}\,(T_{2}\cdot S_{2})_{3} + 
    4\,{\sqrt{{\frac{2}{3}}}}\,(T_{2}\cdot S_{1})_{3} \\
 &  + i\,{\frac{808}{9}}\,{\sqrt{{\frac{2}{3}}}}\,
    ((J_{1}^{2})_{2}\cdot S_{2})_{3}
    + 104\,{\sqrt{{\frac{2}{3}}}}\,((J_{1}^{2})_{2}\cdot S_{1})_{3}\,;
    \vspace{1mm}\\ 
\!\!0 \;\; = & \{\{S_{2},S_{1}\}_{1},S_{2}\}_{3} - 
    i\,{\frac{5}{3}}\,{\sqrt{5}}\,\{\{S_{2},S_{1}\}_{3},S_{1}\}_{3} + 
    i\,{\frac{43}{3}}\,{\sqrt{{\frac{1}{10}}}}\,
     \{\{S_{2},S_{1}\}_{2},S_{1}\}_{3} \\
 &  + i\,24\,{\sqrt{{\frac{1}{5}}}}\,(J_{1}\cdot \{T_{2},S_{1}\}_{3})_{3} + 
    i\,{\frac{8}{3}}\,{\sqrt{10}}\,(J_{1}\cdot \{T_{2},S_{1}\}_{2})_{3} + 
    i\,16\,{\sqrt{{\frac{3}{5}}}}\,(T_{2}\cdot S_{2})_{3} \\
 &  - 32\,{\sqrt{{\frac{3}{5}}}}\,(T_{2}\cdot S_{1})_{3} - 
    i\,{\frac{2528}{3}}\,{\sqrt{{\frac{1}{15}}}}\,
     ((J_{1}^{2})_{2}\cdot S_{2})_{3} - 
    192\,{\sqrt{{\frac{3}{5}}}}\,((J_{1}^{2})_{2}\cdot S_{1})_{3}\,;
    \end{array}  \]
\[  \begin{array}{ll}
    \lefteqn{\!\! J^P=2^+:}  \\
\!\!0 \;\; = & \{\{T_{2},S_{1}\}_{1},S_{2}\}_{2} - 
    {\frac{23}{20}}\,{\sqrt{{\frac{1}{35}}}}\,
     \{\{S_{2},S_{1}\}_{2},T_{2}\}_{2} - 
    {\frac{41}{60}}\,\{\{S_{2},S_{1}\}_{1},T_{2}\}_{2} \\
 &  - i\,{\frac{61}{450}}\,{\sqrt{{\frac{1}{14}}}}\,
     \{\{T_{2},S_{1}\}_{3},S_{1}\}_{2} + 
    i\,{\frac{287}{180}}\,{\sqrt{{\frac{1}{10}}}}\,
     \{\{T_{2},S_{1}\}_{2},S_{1}\}_{2} \\
 &  + i\,{\frac{13}{100}}\,{\sqrt{{\frac{1}{6}}}}\,
    \{\{T_{2},S_{1}\}_{1},S_{1}\}_{2}
    - i\,{\frac{232}{25}}\,{\sqrt{{\frac{2}{7}}}}\,
     (J_{1}\cdot \{S_{2},S_{1}\}_{3})_{2} \\
 &  + i\,{\frac{136}{15}}\,{\sqrt{{\frac{2}{5}}}}\,
     (J_{1}\cdot \{S_{2},S_{1}\}_{2})_{2} - 
    i\,{\frac{51}{25}}\,{\sqrt{6}}\,(J_{1}\cdot \{S_{2},S_{1}\}_{1})_{2} \\
 &  - {\sqrt{{\frac{1}{5}}}}\,(J_{1}\cdot \{S_{1},S_{1}\}_{1})_{2}
    + i\,{\frac{436}{5}}\,{\sqrt{{\frac{2}{105}}}}\,(T_{2}^{2})_{2} + 
    i\,{\frac{171}{5}}\,{\sqrt{{\frac{6}{35}}}}\,(S_{2}^{2})_{2} \\
 &  + {\sqrt{{\frac{1}{15}}}}\,(S_{2}\cdot S_{1})_{2}
    + i\,{\frac{8824}{15}}\,{\sqrt{{\frac{2}{105}}}}\,
     ((J_{1}^{2})_{2}\cdot T_{2})_{2} \\
 &  - i\,{\frac{1232}{15}}\,{\sqrt{{\frac{2}{15}}}}\,
     ((J_{1}^{2})_{0}\cdot T_{2})_{2} 
    + i\,{\frac{2304}{35}}\,{\sqrt{{\frac{6}{5}}}}\,
     ((J_{1}^{2})_{0}\cdot (J_{1}^{2})_{2})_{2}\,;\vspace{1mm}\\ 
\!\!0 \;\; = & \{\{S_{2},S_{1}\}_{3},T_{2}\}_{2} - 
    {\sqrt{{\frac{2}{5}}}}\,\{\{S_{2},S_{1}\}_{2},T_{2}\}_{2} + 
    {\frac{1}{3}}\,{\sqrt{14}}\,\{\{S_{2},S_{1}\}_{1},T_{2}\}_{2} \\
 &  - i\,{\frac{44}{45}}\,\{\{T_{2},S_{1}\}_{3},S_{1}\}_{2} - 
    \,{\frac{i}{9}}\,{\sqrt{{\frac{7}{5}}}}\,
    \{\{T_{2},S_{1}\}_{2},S_{1}\}_{2} + 
    \,{\frac{i}{5}}\,{\sqrt{{\frac{7}{3}}}}\,
    \{\{T_{2},S_{1}\}_{1},S_{1}\}_{2} \\
 &  + i\,{\frac{232}{45}}\,(J_{1}\cdot \{S_{2},S_{1}\}_{3})_{2} - 
    i\,{\frac{32}{9}}\,{\sqrt{{\frac{7}{5}}}}\,
     (J_{1}\cdot \{S_{2},S_{1}\}_{2})_{2} - 
    i\,{\frac{8}{5}}\,{\sqrt{{\frac{7}{3}}}}\,
     (J_{1}\cdot \{S_{2},S_{1}\}_{1})_{2} \\
 &  - i\,64\,{\sqrt{{\frac{1}{15}}}}\,(T_{2}^{2})_{2} - 
    i\,24\,{\sqrt{{\frac{3}{5}}}}\,(S_{2}^{2})_{2} + 
    {\frac{4}{3}}\,{\sqrt{{\frac{70}{3}}}}\,(S_{2}\cdot S_{1})_{2}
    - i\,{\frac{1216}{3}}\,{\sqrt{{\frac{1}{15}}}}\,
     ((J_{1}^{2})_{2}\cdot T_{2})_{2} \\
 &  + i\,{\frac{128}{3}}\,{\sqrt{{\frac{7}{15}}}}\,
     ((J_{1}^{2})_{0}\cdot T_{2})_{2} - 
    i\,256\,{\sqrt{{\frac{3}{35}}}}\,((J_{1}^{2})_{0}\cdot 
    (J_{1}^{2})_{2})_{2}\,; \qquad\qquad\qquad\qquad\qquad
    \end{array}  \]
\[  \begin{array}{ll}
    \lefteqn{\!\! J^P=2^-:}  \\
    \!\! 0 \;\; \stackrel{(*)}{=} &
    \{\{T_{2},S_{2}\}_{1},T_{2}\}_{2} + 
    i\,{\sqrt{{\frac{10}{21}}}}\,\{\{T_{2},S_{1}\}_{2},T_{2}\}_{2} + 
    i\,{\sqrt{{\frac{1}{6}}}}\,\{\{T_{2},S_{1}\}_{1},T_{2}\}_{2} \\
  &  + i\,{\frac{5}{7}}\,{\sqrt{{\frac{3}{2}}}}\,
     \{\{S_{2},S_{1}\}_{1},S_{2}\}_{2} + 
    2\,{\sqrt{{\frac{3}{7}}}}\,\{\{S_{2},S_{1}\}_{3},S_{1}\}_{2} - 
    {\frac{1}{2}}\,{\sqrt{{\frac{5}{3}}}}\,
     \{\{S_{2},S_{1}\}_{2},S_{1}\}_{2} \qquad\qquad\qquad\qquad\qquad \\
 &   + {\frac{1}{7}}\,\{\{S_{2},S_{1}\}_{1},S_{1}\}_{2} - 
    {\frac{i}{14}}\,{\sqrt{{\frac{5}{6}}}}\,
     \{\{S_{1},S_{1}\}_{1},S_{1}\}_{2} - 
    i\,{\frac{14}{5}}\,{\sqrt{{\frac{2}{3}}}}\,
     (J_{1}\cdot \{T_{2},S_{2}\}_{1})_{2} \\
 &   + {\frac{68}{5}}\,{\sqrt{{\frac{3}{7}}}}\,
     (J_{1}\cdot \{T_{2},S_{1}\}_{3})_{2} - 
    {\frac{562}{21}}\,{\sqrt{{\frac{1}{15}}}}\,
     (J_{1}\cdot \{T_{2},S_{1}\}_{2})_{2} - 
    {\frac{32}{7}}\,(J_{1}\cdot \{T_{2},S_{1}\}_{1})_{2} \\
 &   -32\,{\sqrt{{\frac{1}{35}}}}\,(T_{2}\cdot S_{2})_{2} - 
    i\,{\frac{2}{7}}\,{\sqrt{10}}\,(T_{2}\cdot S_{1})_{2} - 
    {\frac{704}{45}}\,{\sqrt{{\frac{1}{35}}}}\,
    ((J_{1}^{2})_{2}\cdot S_{2})_{2} \\
 &   +{\frac{2176}{45}}\,{\sqrt{{\frac{1}{5}}}}\,
    ((J_{1}^{2})_{0}\cdot S_{2})_{2} - 
    i\,{\frac{180}{7}}\,{\sqrt{10}}\,((J_{1}^{2})_{2}\cdot S_{1})_{2}\,;
%    \vspace{1mm}\\
    \end{array} \]
\[  \begin{array}{ll}
\!\! 0 \;\; = &
    \{\{S_{2},S_{1}\}_{1},S_{2}\}_{2} - 
    i\,{\frac{11}{15}}\,{\sqrt{{\frac{7}{2}}}}\,
     \{\{S_{2},S_{1}\}_{3},S_{1}\}_{2} + 
    i\,{\frac{7}{6}}\,{\sqrt{{\frac{1}{10}}}}\,
     \{\{S_{2},S_{1}\}_{2},S_{1}\}_{2} \qquad\qquad\qquad\qquad\qquad\\
 &   + \,{\frac{i}{10}}\,{\sqrt{{\frac{3}{2}}}}\,
     \{\{S_{2},S_{1}\}_{1},S_{1}\}_{2} - 
    {\frac{6}{5}}\,(J_{1}\cdot \{T_{2},S_{2}\}_{1})_{2} - 
    \,{\frac{i}{5}}\,{\sqrt{14}}\,(J_{1}\cdot \{T_{2},S_{1}\}_{3})_{2} \\
 &   +i\,{\frac{7}{3}}\,{\sqrt{{\frac{2}{5}}}}\,
     (J_{1}\cdot \{T_{2},S_{1}\}_{2})_{2} + 
    i\,{\frac{8}{5}}\,{\sqrt{6}}\,(J_{1}\cdot \{T_{2},S_{1}\}_{1})_{2} - 
    i\,2\,{\sqrt{{\frac{42}{5}}}}\,(T_{2}\cdot S_{2})_{2} \\
 &   +i\,{\frac{344}{15}}\,{\sqrt{{\frac{14}{15}}}}\,
     ((J_{1}^{2})_{2}\cdot S_{2})_{2} - 
    i\,{\frac{112}{15}}\,{\sqrt{{\frac{2}{15}}}}\,
     ((J_{1}^{2})_{0}\cdot S_{2})_{2} - 
    96\,{\sqrt{{\frac{3}{5}}}}\,((J_{1}^{2})_{2}\cdot S_{1})_{2}\,;
    \vspace{1mm}\\ 
\!\! 0 \;\; = & \{\{S_{1},S_{1}\}_{1},S_{1}\}_{2} + 
    48\,{\sqrt{{\frac{1}{5}}}}\,(J_{1}\cdot \{T_{2},S_{2}\}_{1})_{2} + 
    i\,4\,{\sqrt{{\frac{14}{5}}}}\,(J_{1}\cdot \{T_{2},S_{1}\}_{3})_{2} \\
 &   -i\,4\,{\sqrt{2}}\,(J_{1}\cdot \{T_{2},S_{1}\}_{2})_{2} + 
    i\,12\,{\sqrt{{\frac{6}{5}}}}\,(J_{1}\cdot \{T_{2},S_{1}\}_{1})_{2} + 
    8\,{\sqrt{3}}\,(T_{2}\cdot S_{1})_{2} \\
 &   -i\,{\frac{64}{5}}\,{\sqrt{{\frac{14}{3}}}}\,
     ((J_{1}^{2})_{2}\cdot S_{2})_{2} - 
    i\,{\frac{448}{5}}\,{\sqrt{{\frac{2}{3}}}}\,
     ((J_{1}^{2})_{0}\cdot S_{2})_{2} - 
    80\,{\sqrt{3}}\,((J_{1}^{2})_{2}\cdot S_{1})_{2}\,;
    \end{array}  \]
\[  \begin{array}{ll}
    \lefteqn{\!\! J^P=1^+:}  \\
\!\!0 \;\; \stackrel{(*)}{=} &
    \{\{T_{2},S_{1}\}_{1},S_{2}\}_{1}
    - {\frac{59}{12}}\,{\sqrt{{\frac{1}{21}}}}\,
     \{\{S_{2},S_{1}\}_{3},T_{2}\}_{1}
    + {\frac{41}{12}}\,{\sqrt{{\frac{1}{15}}}}\,
     \{\{S_{2},S_{1}\}_{2},T_{2}\}_{1}\\
 &  - {\frac{5}{12}}\,\{\{S_{2},S_{1}\}_{1},T_{2}\}_{1}
    + i\,{\frac{29}{12}}\,{\sqrt{{\frac{1}{10}}}}\,
     \{\{T_{2},S_{1}\}_{2},S_{1}\}_{1}
    + i\,{\frac{11}{12}}\,{\sqrt{{\frac{1}{30}}}}\,
     \{\{T_{2},S_{1}\}_{1},S_{1}\}_{1}\\
 &  + {\frac{14}{3}}\,{\sqrt{{\frac{1}{5}}}}\,(J_{1}\cdot 
    \{T_{2},T_{2}\}_{1})_{1} + 
    {\frac{i}{3}}\,{\sqrt{{\frac{2}{5}}}}\,(J_{1}\cdot 
    \{S_{2},S_{1}\}_{2})_{1} + 
    i\,2\,{\sqrt{{\frac{6}{5}}}}\,(J_{1}\cdot \{S_{2},S_{1}\}_{1})_{1}
    \qquad\qquad\qquad\qquad\qquad\\
 &  - {\frac{7}{3}}\,(J_{1}\cdot \{S_{1},S_{1}\}_{1})_{1} + 
    7\,{\sqrt{{\frac{3}{5}}}}\,(S_{2}\cdot S_{1})_{1} + 
    i\,28\,{\sqrt{{\frac{2}{5}}}}\,((J_{1}^{2})_{2}\cdot T_{2})_{1}\,;
    \vspace{1mm}\\
\!\!0 \;\; = & \{\{S_{2},S_{1}\}_{3},T_{2}\}_{1} - 
    {\frac{1}{2}}\,{\sqrt{35}}\,\{\{S_{2},S_{1}\}_{2},T_{2}\}_{1} + 
    {\frac{1}{2}}\,{\sqrt{21}}\,\{\{S_{2},S_{1}\}_{1},T_{2}\}_{1}
    \vspace{1mm}\\
  & - \,{\frac{i}{2}}\,{\sqrt{{\frac{105}{2}}}}\,
     \{\{T_{2},S_{1}\}_{2},S_{1}\}_{1} + 
     \,{\frac{i}{2}}\,{\sqrt{{\frac{35}{2}}}}\,
     \{\{T_{2},S_{1}\}_{1},S_{1}\}_{1} \vspace{1mm}\\
 &  - i\,2\,{\sqrt{210}}\,(J_{1}\cdot \{S_{2},S_{1}\}_{2})_{1}
    - 6\,{\sqrt{35}}\,(S_{2}\cdot S_{1})_{1} - 
    i\,16\,{\sqrt{210}}\,((J_{1}^{2})_{2}\cdot T_{2})_{1}\,;
    \end{array}  \]
\[ \begin{array}{ll}
    \lefteqn{\!\! J^P=1^-:}  \\
\!\!0 \;\; = & \{\{S_{2},S_{1}\}_{1},S_{2}\}_{1} - 
    i\,{\frac{11}{4}}\,{\sqrt{{\frac{1}{10}}}}\,
     \{\{S_{2},S_{1}\}_{2},S_{1}\}_{1} + 
    i\,{\frac{5}{4}}\,{\sqrt{{\frac{5}{6}}}}\,
     \{\{S_{2},S_{1}\}_{1},S_{1}\}_{1} \\
 &  + {\frac{7}{24}}\,\{\{S_{1},S_{1}\}_{1},S_{1}\}_{1} - 
    3\,{\sqrt{{\frac{1}{5}}}}\,(J_{1}\cdot \{T_{2},S_{2}\}_{1})_{1} - 
    12\,{\sqrt{{\frac{1}{5}}}}\,(J_{1}\cdot \{T_{2},S_{2}\}_{0})_{1} \\
 &  + i\,{\frac{14}{3}}\,{\sqrt{10}}\,(J_{1}\cdot \{T_{2},S_{1}\}_{2})_{1} - 
    i\,49\,{\sqrt{{\frac{1}{30}}}}\,(J_{1}\cdot \{T_{2},S_{1}\}_{1})_{1} - 
    i\,3\,{\sqrt{{\frac{2}{5}}}}\,(T_{2}\cdot S_{2})_{1}
    \qquad\qquad\qquad\qquad\qquad\\
 &  + 49\,{\sqrt{{\frac{1}{15}}}}\,(T_{2}\cdot S_{1})_{1} - 
    i\,{\frac{64}{3}}\,{\sqrt{10}}\,((J_{1}^{2})_{2}\cdot S_{2})_{1} + 
    {\frac{322}{3}}\,{\sqrt{{\frac{1}{15}}}}\,
    ((J_{1}^{2})_{2}\cdot S_{1})_{1} \\
 &  - {\frac{196}{3}}\,{\sqrt{{\frac{1}{3}}}}\,
    ((J_{1}^{2})_{0}\cdot S_{1})_{1} \,.
    \end{array} \]
The $\U$--generating relations marked by an asterisk ($*$) result
from induction of $l=2$ relations with the help of the scalar
element $\B_0^{(1)} \in \ga$.
\begin{center}    \underline{$l=4$}   \end{center}
\beas     J^P = 2^-: &\,&  \mbox{~~one $\U$--generating relation induced
                                 from the second of the $l=3$,}
                                 \qquad\\
                      & &  \mbox{~~$J^P = 2^-$ relations with
                                 the help of $\B_0^{(1)}$;}\\
          J^P = 1^+:  & &  \mbox{~~two $\U$--generating relations induced
                                 from the two $l=3$, $J^P = 1^+$} \\
                      & &  \mbox{~~relations with
                                 the help of $\B_0^{(1)}$;}\\
          J^P = 1^-:  & &  \mbox{~~one $\U$--generating relation induced
                                 from the only $l=3$, $J^P = 1^-$} \\
                      & &  \mbox{~~relation with
                               the help of $\B_0^{(1)}$.}
\eeas
\underline{Notation}: The \ir\ tensor variables $(A_{j_1} \cdot 
B_{j_2})_{j_3}$ and $\{ A_{j_1},B_{j_2} \}_{j_3}$ are defined by
\[  \big( A_{j_1} \cdot B_{j_2} \big)_{j_3}  :=
    \left\{  \big( A_{j_1} \cdot B_{j_2} \big)_{j_3,m_3}  \Big| \;
    m_3=-j_3,\dots,+j_3  \right\}  \]
\[  \big\{ A_{j_1},B_{j_2} \big\}_{j_3}  :=
    \left\{ \big\{ A_{j_1},B_{j_2} \big\}_{j_3,m_3}  \Big| \;
    m_3=-j_3,\dots,+j_3  \right\}  \]
with
\[  \big( A_{j_1} \cdot B_{j_2} \big)_{j_3,m_3}  :=
    \langle j_3,m_3 | j_1,m_1;j_2,m_2 \rangle \:
    A_{j_1,m_1} \cdot B_{j_2,m_2}, \]
\[  \big\{ A_{j_1},B_{j_2} \big\}_{j_3,m_3}  :=
    \langle j_3,m_3 | j_1,m_1;j_2,m_2 \rangle  \:
    \big\{ A_{j_1,m_1} , B_{j_2,m_2} \big\}_{\rm PB} . \]
Notice the following symmetry properties:
\[  \big( A_{j_1} \cdot B_{j_2} \big)_{j_3}  =  (-1)^{j_1+j_2-j_3}
    \big( B_{j_2} \cdot A_{j_1} \big)_{j_3} , \]
\[  \big\{ A_{j_1},B_{j_2} \big\}_{j_3}  = - (-1)^{j_1+j_2-j_3}
    \big\{ B_{j_2},A_{j_1} \big\}_{j_3}  .\]
Finally, turning to the action of the basis elements $\B_0^{(l)}$,
$l=1,3,5,\dots$, of the abelian Lie algebra $\ga$ on the generators
$\T_2$, $\S_1$, $\S_2$ of $\U$, the pertinent formulae are exact
copies of the respective formulae valid in the classical Nambu--Goto
theory. For $l=1$ and $l=3$, \ie\ for
\[  \B_0^{(1)}  :=  \frac12  \sum_{j=1}^3  \Z^{+(2)}_{0j0j}  \]
and for
\begin{eqnarray*}   \lefteqn{\!\!
\B_{0}^{(3)} := 12 \, \sum_{j=1}^{3}{{\cal Z}^{+(4)}_{\,000j000j}}\,
 +\,i\,\frac{7}{2}\,\sqrt{\frac{3}{10}}\,\{\{\T_{2},\S_{1}\}_2,\S_2\}_0 \, 
 +\, i\,\frac{15}{2}\,\sqrt{\frac{3}{10}}\{\{\S_2,\S_1,\}_2,\T_2\}_0} \\ &&
 +\,\frac{21}{4\,\sqrt{5}}\,\{\{\T_2,\S_1,\}_1,\S_1\}_0\, 
 +\,i\,7\,\sqrt{\frac{3}{10}}\,\left(\J_1\cdot \{\T_2,\T_2\}_1\right)_0\,
 -\,\frac{21}{2}\,\sqrt{5}\,\left(\J_1\cdot \{\S_2,\S_1\}_1\right)_0 \\ &&
 -\,\frac{28}{\sqrt{5}}\left(\T_2^2\right)_0\, 
 -\,\frac{24}{\sqrt{5}}\,\left(\S_2^2\right)_0
 +\,8\,\sqrt{3}\,\left(\S_1^2\right)_0\,
 -\,\frac{2}{3}\,{\B_0^{(1)}}^2
 +\,\frac{42}{\sqrt{5}}\left(\left(\J_1^2\right)_2\cdot \T_2\right)_0\, ,
\end{eqnarray*}
the latter formulae have been explicitly computed:
\begin{eqnarray*}
  \big\{ \B_{0}^{(1)},\T_{2} \big\}_{2} &=& i\,{\sqrt{6}}\, 
        \left\{ \S_{2},\S_{1} \right\}_{2}\,;\\ 
  \big\{ \B_{0}^{(1)},\S_{1} \big\}_{1} &=& - i\,6\,{\sqrt{{\frac{2}{5}}}}\,
        \left\{ \T_{2},\S_{2} \right\}_{1} + 
        2\,{\sqrt{{\frac{3}{5}}}}\,\left\{\T_{2},\S_{1}\right\}_{1} - 
        24\,{\sqrt{{\frac{3}{5}}}}\,\left(\J_{1}\cdot \S_{2}\right)_{1} \\
  & &   + i\,12\,{\sqrt{2}}\,\left(\J_{1}\cdot \S_{1}\right)_{1}\,;\\ 
  \big\{\B_{0}^{(1)},\S_{2}\big\}_{2} &=& - i\,2\,{\sqrt{{\frac{2}{3}}}}\,
        \left\{\T_{2},\S_{1}\right\}_{2}   - 
        i\,4\,{\sqrt{{\frac{2}{3}}}}\,\left(\J_{1}\cdot \S_{2}\right)_{2} +
        12\,\left(\J_{1}\cdot \S_{1}\right)_{2} \,.
\end{eqnarray*}
The corresponding formulae for $\B_0^{(3)}$ are too lengthy, however,
to be reproduced here.

The $\U$--generating relations formulated as non-identically satisfied
relations in the algebra $\F$ turn into identities when the generators 
of $\U$ are substituted for the generators of $\F$.

Remaining within the limitations of a conventional presentation, the
classical action of the scalar element $\B_0^{(l^\p)}$ of the abelian
Lie algebra $\ga$ on the generators of $\U$ is specified by the
components of the result of the action w.r.t.\ a given basis of
$\U^{l^\p +1}_{j,\pm}$. The choice of this basis is not canonical.
The choice is mainly influenced by aspects of convenience.

On the other hand, the $\U$--generating relations of degree $l$,
spin $j$ and definite parity are quoted in the form of an expansion
in terms of special basis elements of $\F^l_{j,\pm}$. The latter
ones are given as results of Poisson bracket and multiplication
operations involving only subsets of {\it selected} basis elements of
the subspaces $\F^{l^\p}_{j^\p,\pm}$: $l^\p < l$. The said selection 
takes the $\U$--generating relations of degree $l^\p$, spin $j^\p$
and parity plus or minus into account such that the subset of {\it selected}
basis elements of each subspace $\F^{l^\p}_{j^\p,\pm}$, $l^\p < l$,
turns into a basis of the respective subspace $\U^{l^\p}_{j^\p,\pm}$
simply by substituting the generators of $\U$ for the generators of
$\F$. In their turn, the $\U$--generating relations together with
the induced relations, all of them of degree $l$, spin $j$ and
definite parity, also suggest the construction and subsequent
selection of basis elements of $\F^l_{j,\pm}$, the subset of
{\it selected} basis elements virtually providing a basis of the
respective subspace $\U^l_{j,+}$ or $\U^l_{j,-}$: with the help 
of these relations express a maximal number of (as elements of
$\F$) linearly independent $l$--fold rescaled Poisson brackets,
whose entries consist of generators of $\F$, in the form of linear
combinations of the remaining $l$--fold rescaled Poisson brackets
and linearly independent products of {\it selected} basis elements of
$\F^{l^\p}_{j^\p,\pm}$, $l^\p < l$. The said relations can always
be solved for the $l$--fold Poisson brackets by virtue of the
{\it algebraic} independence of the standard invariants \cite{AlgProp}.

A complete set of basis elements of $\F^l_{j,+}$ or $\F^l_{j,-}$ is
given first by a maximal set of linearly independent rescaled Poisson
brackets whose entries consist of the generators of $\F$ and, in addition,
by a maximal set of linearly independent products of arbitrary basis
elements of $\F^{l^\p}_{j^\p,\pm}$, $l^\p < l$, such that the following
rules are satisfied: resulting degree = $l$, resulting spin = $j$ and
resulting parity identical with the definite original parity. The
{\it selected} subset of basis elements of $\F^l_{j,+}$ or $\F^l_{j,-}$
is obtained first by discarding all those $l$--fold
rescaled Poisson brackets which have been expressed with the help
of the generating and the induced relations in the form of the
above mentioned linear combinations and, in addition, keeping
only the linearly independent products of {\it selected} basis
elements of $\F^{l^\p}_{j^\p,\pm}$, $l^\p < l$, satisfying the
above rules for degree, spin and parity.

Note that in none of the above relations there is any explicit 
reference to the typical action $\bf A$, inverse string
tension $\ap$ or mass $\gm$. In fact, all relations are homogeneous
in these parameters such that one can cast them into a form
involving only dimensionless quantities and dimensionless operations.
In the quantum theory this will be carried out explicitly.

\section*{Quantum Theory}

After the above classical preparations I now turn to the main
objective of the present communication: the determination of
the quantum \aoo.
To facilitate the orientation I shall begin with some introductory 
remarks.

One conceivable way to tackle the problem under consideration is
the following: first specify the exact quantum interpretation of
the classical expressions for the generators $\T_2$, $\S_1$, $\S
_2$ and $\B_0^{(l)}$, $l=1,3,5,\dots$, and subsequently from the
resulting commutation relations calculate the quantum \aoo. I
shall {\it not} adopt this strategy in the sequel. In fact, I
shall not construct any concrete non-trivial \r\ at all of the
algebra in question. Instead, along the lines of Ref.\ \cite{Uncov},
I {\it postulate} that the classical generators $(\J_1,)$ $\T_2$,
$\S_1$, $\S_2$ and $\B_0^{(l)}$, $l=1,3,5,\dots$, possess faithful
quantum counterparts $(\JH_1,)$ $\TH_2$, $\SH_1$, $\SH_2$ and
$\BH_0^{(l)}$, $l=1,3,5,\dots$, respectively, which carry exactly
the same dimensions and have the same covariance properties under
rotations, reflections and star operation as their classical
partners into which they turn in the classical limit.

Further, I {\it postulate} that there is a one to one correspondence
between the generating relations of the Poisson algebra with generators
$\J_1$, $\T_2$, $\S_1$, $\S_2$ and $\B_0^{(l)}$, $l=1,3,5,\dots$,
and the generating relations of the associative non-commutative 
algebra with generators $\JH_1$, $\TH_2$, $\SH_1$, $\SH_2$ and
$\BH_0^{(l)}$, $l=1,3,5,\dots$, and that there is a structural
similarity (to be explained below) between the classical and the
quantum generating relations. Consistency of these requirements
-- in particular under commutator operations -- should remove
remaining ambiguities.

Once a relevant part of the quantum algebra $\ggh$ (or, equivalently,
$\ghh$) has been determined, one can indeed proceed to specify the
exact quantum interpretation of the classical expressions for $\T_2$,
$\S_1$, $\S_2$ and more and more scalar elements $\B_0^{(l)}$ by
using the previously established relations and commutator actions as a guide
line. Moreover, by invoking the technique of induced \r s \cite{Ni}
one can finally arrive at a faithful \r\ of the Poincar\'e invariant
version of the quantum algebra of observable symmetry transformations.
However, this aspect of the quantization
program has not been pursued in any detail, yet, and will not be
pursued here. It will be the subject of future investigations.

Now I set about determining a relevant part of the quantum algebra
of internal \obs s built from the left movers. From the very beginning
I shall {\it assume} that the full quantum \aoo\ is Poincar\'e
invariant and that the components $\PH_\mu$ of the energy momentum
operator continue to be central elements of the algebra of internal
invariant charges. As before, this assumption permits to treat the 
components $\PH_\mu$ as concrete c--numbers $p_\mu$ and to pass to 
the rest frame of $p_\mu$:
\[  \PH_\mu  =  \gm \, \delta_\mu^0  \, \id . \]
The associative non-commutative algebra of \obs s obtained in 
this way is denoted by $\ghh$. It is {\it postulated} that $\ghh$
is generated by elements $\JH_1$, $\TH_2$, $\SH_1$, $\SH_2$, and
$\BH_0^{(l)}$, $l=1,3,5,\dots$. Further, since in the quantum 
theory the typical action $\bf A$ is given by $\hslash$
(= Planck's constant divided by $2\pi$), it is {\it postulated}
that the generators $\JH_1$, $\TH_2$, $\SH_1$, $\SH_2$ and
$\BH_0^{(l)}$ scale like dimensionless scale invariant operators
$\JHQ_1$, $\THQ_2$, $\SHQ_1$, $\SHQ_2$ and $\BHQ_0^{(l)}$ times 
$(\hslash/2\pi\ap\gm^2) \gm^2$, $(\hslash/2\pi\ap\gm^2)^2 \gm^4$,
$(\hslash/2\pi\ap\gm^2)^2 \gm^4$, $(\hslash/2\pi\ap\gm^2)^2 \gm^4$
and $(\hslash/2\pi\ap\gm^2)^{l+1} \gm^{2(l+1)}$, respectively.
In particular, each one of these generators of the algebra $\ghh$
-- in the original as well as in the dimensionless form -- is
affiliated with a definite non-negative integer degree $l$ or
-- equivalently -- with a definite positive order in Planck's
constant. This order is given by the respective power in the
dimensionless constant $\hslash/(2\pi\ap\gm^2)$. It agrees
with the ``grade'' $(l+1)$ of the respective classical partner.
To lowest order in Planck's constant, the Poisson bracket is
replaced by the commutator with the help of the following substitution:
\[  \Big\{ \;,\; \Big\}^{\rm PB}  \longrightarrow  \frac1{i\hslash}
    \Big[ \;,\; \Big]  \]
or, equivalently,
\[  \Big\{ \;,\; \Big\}_{\rm PB}  \longrightarrow  \frac1{i\left(
    \frac{\hslash}{2\pi\ap}\right)}  \Big[ \;,\; \Big].  \]
Requiring structural similarity with the Poisson algebra $\gh$, the
associative algebra $\ghh$ is {\it postulated} to be the enveloping
algebra of $\ggh := \gah \ltimes \UH$, $\UH$ being an associative
algebra. The product in the enveloping algebra and the product in
$\UH$ are to be identified. The algebra $\UH$ is generated by $\JHQ_1$,
$\THQ_2$, $\SHQ_1$, $\SHQ_2$. The first summand $\gah$ is an abelian
commutator Lie algebra generated by the scalar, (w.r.t.\
$*$--operation) self conjugate basis elements $\BHQ_0^{(l)}$, $l=
1,3,5,\dots$:
\[  \Big[ \BHQ_0^{(l_1)} , \BHQ_0^{(l_2)} \Big] = 0  \qquad\qquad
    \qquad \mbox{for}
    \:\: l_1,l_2 \in \{ 1,3,5,\dots \},   \]
which act as derivations on $\UH$.

$\UH$ will be presented as a two-fold quotient of the free associative
algebra $\FH_0$ with abstract generators $\SJH_1$, $\STH_2$, $\SSH_1$,
$\SSH_2$ of the same dimensions and affiliated with the same orders
in Planck's constant as $\JH_1$, $\TH_2$, $\SH_1$, $\SH_2$,
respectively. The order in $\hslash$ provides a grading of $\FH_0$
(w.r.t.\ multiplication).

In complete congruency with the classical situation, the first 
quotient of $\FH_0$ takes care of the commutation relations of
the components of $\JH_1$ among each other, of the commutation
relations of the components of $\JH_1$ with the components of
$\TH_1$, $\SH_1$ and $\SH_2$ and of the relations among each of
the \ir\ tensor operators under consideration resulting from the
$*$--operation and parity transformation. The result of the above
identifications is an associative algebra $\FH$ with generators
$\HJ_1$, $\HT_2$, $\HS_1$, $\HS_2$ of dimensions $\gm^2$,
$\gm^4$, $\gm^4$, $\gm^4$ and of orders in Planck's constant 1,
2, 2, 2, respectively. The order in $\hslash$ provides an
$\N$--grading of $\FH$ w.r.t.\ multiplication.

Given any concrete choice of the degree $l$, $l \in \{ 0,1,2,\dots\}$
(or, equivalently, of the order $(l+1)$ in Planck's constant), of the
spin $j$, $j \in \{ 0,1,2,\dots,l+1 \}$, and
of the parity, translate the basis of the subspace $\F^l_{j,\pm}$
of $\F$ constructed before into a basis of the homogeneous
subspace $\FH^l_{j,\pm}$ of $\FH$ according to the following
rules:

\begin{itemize}
\item[{\it i})]
replace each (simple) rescaled Poisson bracket $\{ \,,\, \}_{j^\p}
:= 2\pi\ap \{ \,,\, \}_{j^\p}^{\rm PB}$ of the classical \ir\
tensor variables by $(i\hslash/2\pi\ap)^{-1}$ times the commutator
$[ \,,\, ]_{j^\p}$ of their respective counterparts in $\FH$ without
changing the original succession of brackets inside an iterated
bracket;

\item[{\it ii})]
replace each (simple) product $(\,\cdot\,)_{j^\p}$ of the classical
\ir\ tensor variables by $1/2$ times the anticommutator $\{\,,\,\}
_{j^\p}$ of their respective counterparts in $\FH$ without changing 
the original succession of brackets inside an iterated bracket.

\end{itemize}
(These rules are suggested by the relative simplicity of the
recoupling formulae for (mixed) commutators and anticommutators
of three irreducible tensor operators (see further below).)

\vspace{2mm}
Similarly, for every fixed degree $l$, spin $j$ and parity for
which all classical relations -- $\U$--generating relations as 
well as induced relations -- are explicitly known and for which
the entire set of classical relations of lesser degree is also
known in explicit form, translate the subset of selected basis
elements of $\F^l_{j,\pm}$ constructed before into a corresponding
subset of selected basis elements of $\FH^l_{j,\pm}$. Finally,
one more substitution ($\JHQ_1$, $\THQ_2$, $\SHQ_1$, $\SHQ_2$
for $\HJ_1$, $\HT_2$, $\HS_1$, $\HS_2$, respectively) turns
the selected basis elements of $\FH^l_{j,\pm}$ into elements of
the isotypical component $\UH^l_{j,\pm}$ of $\UH^l \subset \UH$.
Under the provision that the
consistency postulate is rigorously satisfied for all cycles of
the deformation routine described below, the latter elements form
a (linear) basis of the subspace $\UH^l_{j,\pm}$. The algebra
$\UH$ is still the direct sum of all such vector spaces $\UH^l_{j,\pm}$
though this sum does not correspond any more to an $\N_0$--grading
w.r.t.\ the commutator operation. Instead it corresponds to a
filtration.

The second quotient of $\FH_0$, \ie\ the first and only quotient
of the graded algebra $\FH$, is performed w.r.t.\ an ideal $\IH$
whose generators are deduced from the (quantum) $\UH$--generating
relations. In their turn the $\UH$--generating relations are
obtained, on the one hand, from the classical {\it truly independent}
$\U$--generating relations (see above) and the classical formulae for
the respective actions of the basis elements $\B_0^{(l)}$ of the
abelian algebra $\ga$ on the generators of $\U$ by a fairly
conservative deformation of these classical relations and formulae
and, on the other hand, from the postulated commutativity of the
quantum actions of different $\BHQ^{(l)}_0$'s. As a result of this
deformation also the quantum action of the basis elements $\BHQ_0^
{(l)}$ of the abelian algebra $\gah$ on the generators of $\UH =
\FH \big/ \IH$ is given, albeit -- to begin with -- in parametrized
form.

Now I shall describe the deformation procedure. Consider the 
$\U$--generating relations {\it without} an asterisk, \ie\ the
{\it truly independent} generating relations of the classical Nambu--Goto
theory. Since there are no classical $\U$--generating relations at
all for degree $l=0$ and $l=1$, and thus by the consistency {\it
postulate} there will never be any quantum relations affiliated
with these degrees, the corresponding subspaces $\UH^l_{j,\pm}$
of $\UH$ with $l=0$ and $l=1$ can be identified with the respective
subspaces $\FH^l_{j,\pm}$ of $\FH$:
\[  \FH^0_1  =  \FH^0_{1,+} ;  \qquad\qquad  \UH^0_1  =  \UH^0_{1,+} \]
where $\FH^0_{1,+}$ and $\UH^0_{1,+}$ are the linear spans of 
$\HJ_{1,m}$ and $\JHQ_{1,m}$, respectively;
\[  \FH^1_+ = \FH^1_{2,+} , \qquad
    \FH^1_- = \FH^1_{1,-} \oplus \FH^1_{2,-} ;\qquad
    \UH^1_+ = \UH^1_{2,+} , \qquad
    \UH^1_- = \UH^1_{1,-} \oplus \UH^1_{2,-}  \]
where $\FH^1_{2,+}$ and $\UH^1_{2,+}$ are the linear spans of 
$\HT_{2,m}$ and $\THQ_{2,m}$, respectively, and -- with $i$ equal
to 1 or 2 -- $\FH^1_{i,-}$ and $\UH^1_{i,-}$ the linear spans of
$\HS_{i,m}$ and $\SHQ_{i,m}$, respectively.

The first non-trivial cycle of the deformation routine begins with
the deformation of the $\U$--generating relations of lowest degree,
\ie\ of degree 2, and the deformation of the formulae specifying
the action of the scalar basis element $\B_0^{(1)}$ on the generators
of $\U$. Having completed the first non-trivial cycle of the
deformation routine, advance degree by degree: from $l$ to $(l+1)$
as long as the consistency requirements are met and as far as the 
classical preparations permit.

A full cycle of the deformation routine, using the constructs of 
the previous cycles and/or the displayed features of the $l=0$ 
and $l=1$ subspaces, consists of the following successive steps
(the first two of which are already familiar):

In every one of the truly independent $\U$--generating relations
of degree $(l+1)$
\begin{itemize}

\item[{\it i})]
replace each (simple) rescaled Poisson bracket $\{ \,,\, \}_j := 
2\pi\ap \{ \,,\, \}_j^{\rm PB}$ of the classical \ir\ tensor variables
by $(i\hslash/2\pi\ap)^{-1}$ times the commutator $[ \,,\, ]_j$ of 
their respective counterparts in $\FH$ without changing the original 
succession of brackets inside an iterated bracket;

\item[{\it ii})]
replace each (simple) product $(\,\cdot\,)_j$ of the classical \ir\ 
tensor variables by $1/2$ times the anticommutator $\{\,,\,\}_j$ 
of their respective quantum counterparts (without changing the 
original succession of brackets inside an iterated bracket);

\item[{\it iii})]
add the most general admissible quantum correction, \ie\ the most
general parametrized linear combination of the basis elements of
$\UH^{l^\p}_{j,\pm}$ with $0\le l^\p \le l$ compatible with the
mass dimension, the order $(l+2)$ in Planck's constant, the spin,
the parity and the reality property (under $*$--operation) of
the initial classical relation.

\item[{\it iv})]
Divide the relation so obtained by $(\hslash/2\pi\ap)^{l+2}$
to arrive at a dimensionless parametrized relation involving
the dimensionless generators $\HJQ_1$, $\HTQ_2$, $\HSQ_1$,
$\HSQ_2$, dimensionless operations and dimensionless parameters.

\end{itemize}
At this point the deformed versions of a {\it maximal} set of
{\it linearly independent} relations of degree $(l+1)$ are available, 
though partially in parametrized form only. The deformed versions
of the complement of the truly independent $\U$--generating relations
have already been obtained (partially in parametrized form) in the
course of the preceding cycle by way of induction employing among
other things the possibly parametrized commutator action of scalar
basis elements $\BHQ_0^{(l^\p)}$ with $l^\p < l$. When all the 
parameters are finally fixed, each relation of the maximal set 
under discussion must turn into an identity in $\UH$.

After this digression, I shall resume the description of the
successive steps making up a full cycle:
\begin{itemize}

\item[{\it v})]
In case that $(l+1)$ is even, apply the above steps $i$) -- $iii$)
to the formulae speci- fying the classical action of the scalar
basis element $\B_0^{(l)} \in \ga$ on the generators of $\U$
to obtain the corresponding parametrized quantum action of the
scalar basis element $\BHQ_0^{(l)} \in \gah$ on the generators
$\THQ_2$, $\SHQ_1$ and $\SHQ_2$ of $\UH$.

\item[{\it vi})]
Produce and analyze the set of all $\UH$--relations of degree
$(l+2)$ induced, on the one hand, from the partially parametrized
$\UH$--generating relations of degree $(l+1)$ and less in all
possible, {\it a priori} independent ways including, in particular,
the possibly parametrized commutator action of the basis elements
of $\gah$: $\BHQ_0^{(l^\p)}$, $l^\p=1,3,5,\dots$, $l^\p \le l$, and,
on the other hand, 
from the postulated commutativity of the possibly parametrized
commutator actions of different basis elements $\BHQ_0^{(l^\p)}$
and $\BHQ_0^{(l^{\p\p})}$ with $l^\p + l^{\p\p} \le l+1$ ({\it cf}
the comments below).
\\[1mm]
Insisting on the absence of any quantum relation which does
not correspond unambiguously to a valid classical relation
-- this is the decisive consistency {\it postulate} -- derive
as many independent restrictions as possible for the parameters
involved and resolve these restrictions by assigning numerical
values to some or all of the parameters and by expressing the
remaining parameters with the help of as few (old or new)
parameters as possible. The restrictions are algebraic. Without
loss of generality the parameters both old and new may be chosen
to be real.

\item[{\it vii})]
Deduce the respective generators (modulo parameter fixing) of
$\IH$ from the possibly still parametrized $\UH$--generating
relations of degree $(l+1)$.

\item[{\it viii})]
Start a new cycle of the deformation routine with the degree
raised by one and determine -- if possible -- in the course 
of the due consistency checks, besides the restrictions for
the additional parameters to be introduced,  the numerical
values of the parameters of the previous quantum relations
and quantum actions or, at least, reduce the number of the
free parameters involved in them.

\end{itemize}
Here is an example of the quantum deformation of a classical
$\U$--generating relation (without an asterisk) of degree
$(l+1)$: $(l+1)=2$, $J^P = 1^+$, the classical relation is
\[  \begin{array}{rcl}
    0 &=& i\,\{S_2,S_2\}_1\,
    -\, \sqrt{\frac{2}{3}}\,\{S_2,S_1\}_1\,
    +\, \frac{i}{6}\,\sqrt{5}\, \{S_1,S_1\}_1 \\
&&  -\, 16\, \sqrt{\frac{2}{3}}\, (J_1 \cdot T_2)_1\,
    -\,32\,\sqrt{\frac{2}{15}}\,(J_1 \cdot (J_1^2)_0)_1\,;
    \end{array} \]
its parametrized $\UH$--generating counterpart of order three in
Planck's constant is given by
\[   \begin{array}{rcl}
     0 &=& [\HSQ_2,\HSQ_2]_1\,
     +\, i\, \sqrt{\frac{2}{3}}\,[\HSQ_2,\HSQ_1]_1\,
     +\, \frac{1}{6}\,\sqrt{5}\, [\HSQ_1,\HSQ_1]_1\\&&
     -\, 8\, \sqrt{\frac{2}{3}}\, \{\HJQ_1,\HTQ_2\}_1\,
     -\,16\,\sqrt{\frac{2}{15}}\,\{\HJQ_1,(\HJQ_1^{\,2})_0\}_1\,
     +\, f\, \sqrt{10}\, \HJQ_1 \,.
     \end{array}  \]
The corresponding parametrized quantum generator of $\IH$ is:
\[   \begin{array}{rcl}
     \quad && [\HSQ_2,\HSQ_2]_1\,
     +\, i\, \sqrt{\frac{2}{3}}\,[\HSQ_2,\HSQ_1]_1\,
     +\, \frac{1}{6}\,\sqrt{5}\, [\HSQ_1,\HSQ_1]_1\\&&
     -\, 8\, \sqrt{\frac{2}{3}}\, \{\HJQ_1,\HTQ_2\}_1\,
     -\,16\,\sqrt{\frac{2}{15}}\,\{\HJQ_1,(\HJQ_1^{\,2})_0\}_1\,
     +\, f\, \sqrt{10}\, \HJQ_1\, .
     \end{array}  \]
The symbol $f$ employed above denotes a real valued parameter, which
in its initial form, $F=\big( \frac\hslash{2\pi\ap} \big)^2 f$, is
of order two in Planck's constant ({\it cf} comments below).

An example of the quantum deformation of the classical formulae
of degree $(l+1)$ describing the classical action of the basis
element $\B_0^{(l)}$ of $\ga$ on $\U$ is provided by $(l+1) = 2$,
$J^P = 1^-$,
\[   \begin{array}{rcl}
     i\,\{\B_0^{(1)},\S_1\}_1 &=& 6\, \sqrt{\frac{2}{5}}\, \{\T_2,\S_2\}_1\,
     +\, i\, 2\, \sqrt{\frac{3}{5}}\, \{\T_2,\S_1\}_1 \,
     -\, i\, 24\,\sqrt{\frac{3}{5}}\, (\J_1 \cdot \S_2)_1 \\&&
     -\, 12\, \sqrt{2}\, (\J_1 \cdot \S_1)_1\, .
     \end{array} \]
The quantum action of $\BHQ_0^{(1)}$ on $\SHQ_1$ is given by the
parametrized relation of order three in Planck's constant
\[   \begin{array}{rcl}
     [\BHQ_0^{(1)},\SHQ_1]_1 &=&
     -\, i\, 6\, \sqrt{\frac{2}{5}}\, [\THQ_2,\SHQ_2]_1\,
     +\, 2\, \sqrt{\frac{3}{5}}\, [\THQ_2,\SHQ_1]_1\,
     -\, i\, 12\,\sqrt{\frac{3}{5}}\, \{\JHQ_1,\SHQ_2\}_1\\&&
     -\, 6\, \sqrt{2}\, \{\JHQ_1,\SHQ_1\}_1\,
     +\, i\, d\, \SHQ_1\, .
     \end{array}  \]
Here, the symbol $d$ denotes a real valued parameter, which in its
initial form, $D=\frac\hslash{2\pi\ap}\, d$, is of order one in
Planck's constant.
\\[2mm]
A few comments upon the deformation procedure may be helpful.
\\[2mm]
{\it Concerning step iii}):  Each coefficient of the parametrized
linear combination
under discussion carries an explicit factor $(\hslash/2\pi\ap
\gm^2)^{l-l^\p}$, $l^\p < l$, and an explicit mass dimension
in order to balance the difference between the order in Planck's
constant and the mass dimension of the respective basis element
of $\UH^{l^\p}_{j,\pm}$ on the one hand and the common order in
$\hslash$ and common mass dimension of the terms arising from the 
classical relation in question through operations $i$) and $ii$)
on the other hand.
\\[2mm]
{\it Concerning steps iv}) -- $vi$): It would be inconsistent with
insertion of numerical values to assign to the parameters in
question also in their dimensionless form definite respective
degrees or -- equivalently -- orders in Planck's constant.
Hence the individual terms of a parametrized or numerically
specified $\UH$--generating relation in dimensionless form
are no longer uniform w.r.t.\ the degree. The quantum
counterparts of the respective terms of the classical relation
still have the largest degree present in common, the smaller
degrees being carried by components which arise from the 
individual constituents of the quantum correction.
\\
Nevertheless, also in this form the degree $l$ or -- 
equivalently -- the order $(l+1)$ in Planck's constant of the
relation before division can be read off and can be used as a
label for the relation after division. The same goes for the
dimensionless relations parametrizing the quantum action of
the basis elements $\BHQ_0^{(l^\p)}$ of $\gah$ on $\UH$, for 
the remaining dimensionless $\UH$--generating relations
induced with the help of the elements $\BHQ_0^{(l^\p)}$ and
for the generators deduced from the dimensionless $\UH$--generating
relations.
\\
Of course, the degree $l$ no longer endows the associative
algebras $\FH$ and $\UH$, and their respective subspaces $\FH
_{j,\pm}$ and $\UH_{j,\pm}$, with an $\N_0$--grading w.r.t.\ the
commutator operation. It endows them with a filtration and
possibly -- as will be discussed later -- with a $\Zet_2$--grading,
instead. Moreover, in analogy to the classical situation, the
degree behaves differently under (rescaled) commutator and
under anticommutator operation. Consequently, no well-defined
degree can be assigned to naive products of factors which do
not commute with each other in the spin channel under consideration.
\\[2mm]
{\it Concerning steps vi}) -- $viii$): Associativity of the algebra
w.r.t.\ the underlying multiplication is frequently taken into
account by arranging the entries $A_{j_1}$, $B_{j_2}$, $C_{j_3}$
of iterated commutators, anticommutators or mixtures thereof in
a standard order with the help of the following three formulae.
The schemes $\left\{ {j_1 \atop l_1}\,{j_2 \atop l_2}\,{j_3 \atop 
l_3} \right\}$ denote the corresponding 6$j$--symbols \cite{Metrop}.
\beas  \lefteqn{ \Big[ \Big[ A_{j_1},B_{j_2} \Big]_l , C_{j_3} \Big]_L 
                 = \sum_k \;(-1)^{k+j_2+j_3} \sqrt{(2l+1)(2k+1)} \quad
                 \times \qquad\qquad\qquad\qquad } \\
          & &    \quad\times \: \Bigg( (-1)^l \left\{ {j_2 \atop j_3}\,
                 {j_1 \atop L}\,{l \atop k} \right\}  
                 \Big[ \Big[ A_{j_1},C_{j_3} \Big]_k , B_{j_2} \Big]_L -
                 \left\{ {j_1 \atop j_3}\,{j_2 \atop L}\,{l \atop k} 
                 \right\} \Big[ \Big[ B_{j_2},C_{j_3} \Big]_k , A_{j_1} 
                 \Big]_L \Bigg) , \\
       \lefteqn{ \Big[ \Big\{ A_{j_1},B_{j_2} \Big\}_l , C_{j_3} \Big]_L 
                 = \sum_k \;(-1)^{k+j_2+j_3} \sqrt{(2l+1)(2k+1)} \quad
                 \times \qquad\qquad } \\ 
          & &    \quad\times \: \Bigg( (-1)^l \left\{ {j_2 \atop j_3}\,
                 {j_1 \atop L}\,{l \atop k} \right\}  
                 \Big\{ \Big[ A_{j_1},C_{j_3} \Big]_k , B_{j_2} \Big\}_L +
                 \left\{ {j_1 \atop j_3}\,{j_2 \atop L}\,{l \atop k}
                 \right\} \Big\{ \Big[ B_{j_2},C_{j_3} \Big]_k , A_{j_1} 
                 \Big\}_L \Bigg) , \\
       \lefteqn{ \Big\{ \Big\{ A_{j_1},B_{j_2} \Big\}_l , C_{j_3} \Big\}_L 
                 = \sum_k \;(-1)^{k+j_2+j_3} \sqrt{(2l+1)(2k+1)} \quad
                 \times \qquad\qquad } \\ 
          & &    \quad\times \: \Bigg( (-1)^{l+1} \left\{ {j_2 \atop j_3}\,
                 {j_1 \atop L}\,{l \atop k} \right\}  
                 \Big[ \Big[ A_{j_1},C_{j_3} \Big]_k , B_{j_2} \Big]_L +
                 \left\{ {j_1 \atop j_3}\,{j_2 \atop L}\,{l \atop k} 
                 \right\} \Big\{ \Big\{ B_{j_2},C_{j_3} \Big\}_k , A_{j_1} 
                 \Big\}_L \Bigg) . \eeas
There is a momentous  difference between the first two formulae
on the one hand and the third formula on the other hand: whereas
all terms of the first two formulae involve a common number of
commutator operations and a common number of anticommutator
operations, the first term inside the parenthesis on the right
hand side of the third formula employs two commutator operations
instead of the two anticommutator operations of the remaining 
terms of this formula. Thus, in general, elements of the algebra
$\ghh$ in their dimensionless form, homogeneous w.r.t.\ the order
in Planck's constant, split off elements of lesser degree when
they are subjected to manipulations involving the third formula.
This does not happen when they are subjected to manipulations
involving either one of the first two formulae.
\\
Of course, there exist also valid classical versions of the above
three formulae. They are obtained by substituting Poisson brackets
for the commutator brackets in the first two formulae and by
dropping the first term inside the parenthesis on the right hand
side of the third formula. In fact, these classical versions have
been used over and over again beforehand in the course of the 
classical considerations.
\\[2mm]
{\it Concerning step vi}): Suppose that one has confined oneself
to the subspace $\left( \bigoplus_{l=1}^{k-1} \ga^l \right)$ $\oplus
\left( \bigoplus_{l=0}^k \F^l \right)$ of $\gh$ for any fixed $k
\ge 1$ and that one has determined the set of all classical relations
and actions available under this confinement. Then the consistency of
the Poisson algebra $\gh$ guarantees that the relations, obtained 
from the above set by all possible moves of induction into the
subspace $\bigoplus_{l=0}^{k+1} \F^l$, never lash back at the
subspace $\left( \bigoplus_{l=1}^{k-1} \ga^l \right) \oplus \left(
\bigoplus_{l=0}^k \F^l \right)$ by imposing (algebraic) relations
for the elements of the latter which do not turn into identities
when the initial relations and actions are taken into account.
\\
In the quantum theory based on the deformation of the Poisson
algebra, the consistency of the latter guarantees the following
much weaker property of the corresponding set of partially
parametrized deformed relations and actions: in case that induction
of these relations and actions produces degree$(k+1)$--relations
involving as their only respective contribution with degree $(k+1)$
a linear combination of anticommutators, then each such linear 
combination in its entirety turns into a sum of terms with lesser
degree when the initial deformed relations and actions are taken
into account {\it without} further adjustment of their parameters.
\\
The consistency {\it postulate} requires that the resulting relations
with degree $\le k$ turn into identities when appropriate numerical
values are assigned once and for all to some of the parameters or
to all of them and when the initial deformed relations and actions
are taken into consideration with these numerical values for their
respective parameters. There is no escape from an overdetermined 
system of equations for the parameters that would not violate the
consistency postulate.

\vspace{2mm}

So far the deformation routine has been carried out for the cycles
of degree two and three:

In the course of the cycle of degree two, the first non-trivial
cycle, six free real parameters were introduced: in their initial
form five of first order and one of second order in Planck's
constant. Step $vi$) of the deformation routine did not yield any
restriction on these parameters.

In the course of the cycle of degree three, twenty-nine additional
free real parameters were introduced: twenty one initially of first,
seven initially of second and one initially of third order in
Planck's constant. After a lot of processing, step $vi$) of the
deformation routine furnished a system of linear and quadratic
restrictions for the thirty-five parameters. The system received
multiple contributions from each of the final spin--parity channels
$J^P = 0^\pm,\,2^\pm,\,3^\pm$ and $4^\pm$. At first sight the
system looked hopelessly overdetermined. From it without much
difficulty a system of twenty-six independent homogeneous linear
equations for twenty-six parameters was split off, all of the
latter in their initial form of first order in Planck's constant.
Necessarily these twenty-six parameters had to be set equal to
zero. These insertions immediately implied that also the parameter
initially of third order in Planck's constant had to be set equal
to zero. At this point the original system of linear and quadratic
equations was reduced to a system of linear inhomogeneous equations
for the remaining eight parameters, all of which initially of
second order in Planck's constant. (Multiple) contributions to
this system came from the final spin--parity channels $J^P = 
0^-,\,2^\pm$ and $3^\pm$. Without prior knowledge this system
still looked grossly overdetermined. At the end of the day,
however, it turned out that the latter system was not overdetermined
at all. On the contrary, it furnished only seven independent
linear inhomogeneous equations for the remaining eight parameters.
The numerical value of one of these parameters appearing in the
basic parametrization of the quantum deformation was fixed as a
rational number, to wit $16/5$. The rest of the equations was
used to express six of the other seven parameters in terms of
the only parameter initially of second order in Planck's constant
which made its appearance already in the course of the cycle of 
degree two. The coefficients of the respective expressions are
rational numbers.

\vspace{2mm}

Here are the basic data of the quantum \aoo\ of the Nambu--Goto
theory as they have been obtained so far. They are quoted in
condensed form in terms of those ``fundamental'' generators of
$\IH$ corresponding to {\it truly independent} $\UH$--generating
relations and of the quantum action of $\BHQ_0^{(1)}$ on $\UH$.
They still require the fixing and subsequent substitution of
the numerical value of the residual parameter $f$.

\begin{center}  \underline{Fundamental generators of $\IH$ of
                           degree two:}
\end{center}
\[  \begin{array}{ll}
\!\!J^P=4^-:\qquad & [\HTQ_{2},\HSQ_{2}]_{4}\,; \vspace{1mm}\\
\!\!J^P=3^+: & [\HTQ_{2},\HTQ_{2}]_{3} 
           + i\,[\HSQ_{2},\HSQ_{1}]_{3} 
           + 16\,(\HJQ_{1}^{\,3})_{3}\,; \vspace{1mm}\\
         & [\HSQ_{2},\HSQ_{2}]_{3} 
           - i\,2\,[\HSQ_{2},\HSQ_{1}]_{3} 
           - 4\,\{\HJQ_{1},\HTQ_{2}\}_{3} 
           - 48\,(\HJQ_{1}^{\,3})_{3}\,;
           \qquad\qquad\qquad\qquad\qquad\qquad\qquad\qquad\qquad
           \vspace{1mm}\\
\!\!J^P=3^-: & [\HTQ_{2},\HSQ_{2}]_{3} 
           - i\,[\HTQ_{2},\HSQ_{1}]_{3} 
           + 4\,\{\HJQ_{1},\HSQ_{2}\}_{3}\,; \vspace{1mm}\\
\!\!J^P=2^-: & [\HTQ_{2},\HSQ_{2}]_{2} 
           + {\frac{i}{3}}\,{\sqrt{{\frac{7}{2}}}}\, [\HTQ_{2},\HSQ_{1}]_{2}
           - {\frac{2}{3}}\,{\sqrt{14}}\,\{\HJQ_{1},\HSQ_{2}\}_{2}\,;\\ 
\!\!J^P=1^+: & [\HSQ_{2},\HSQ_{2}]_{1} 
           + i\,{\sqrt{{\frac{2}{3}}}}\,[\HSQ_{2},\HSQ_{1}]_{1} 
           + {\frac{1}{6}}\,{\sqrt{5}}\,[\HSQ_{1},\HSQ_{1}]_{1}\\& 
           - 8\,{\sqrt{{\frac{2}{3}}}}\,\{\HJQ_{1},\HTQ_{2}\}_{1} 
           - 16\,{\sqrt{{\frac{2}{15}}}}\,
             \{\HJQ_{1},(\HJQ_{1}^{\,2})_{0}\}_{1} 
           + f\, {\sqrt{10}}\,\HJQ_{1} 
   \end{array}  \]
\newpage
\begin{center}  \underline{Fundamental generators of $\IH$ of
                           degree three:}
\end{center}
\[  \begin{array}{ll}
    \lefteqn{\!\!J^P=5^-:} \vspace{1mm}\\
&   [[\HSQ_{2},\HSQ_{1}]_{3},\HSQ_{2}]_{5}\,;
    \qquad\qquad\qquad\qquad\qquad\qquad
    \qquad\qquad\qquad\qquad\qquad\qquad
    \qquad\qquad\qquad\qquad\qquad\qquad
    \end{array}  \]
\[  \begin{array}{ll}
    \lefteqn{\!\! J^P=4^+:}\\
&  [[\HSQ_{2},\HSQ_{1}]_{2},\HTQ_{2}]_{4} + 
   i\,{\frac{4}{9}}\,{\sqrt{{\frac{2}{3}}}}\,
   [[\HTQ_{2},\HSQ_{1}]_{3},\HSQ_{1}]_{4} - 
   {\frac{20}{9}}\,{\sqrt{{\frac{2}{3}}}}\,
   \{\HJQ_{1},[\HSQ_{2},\HSQ_{1}]_{3}\}_{4} - 
   i\,{\frac{16}{3}}\,{\sqrt{{\frac{2}{3}}}}\,(\HTQ_{2}^{\,2})_{4}
   \qquad\qquad\qquad\qquad\qquad \\
&  + i\,4\,{\sqrt{{\frac{2}{3}}}}\,(\HSQ_{2}^{\,2})_{4} + 
   i\,{\frac{32}{3}}\,{\sqrt{{\frac{2}{3}}}}\,
   \{(\HJQ_{1}^{\,2})_{2},\HTQ_{2}\}_{4} + 
   i\,{\frac{128}{3}}\,{\sqrt{{\frac{2}{3}}}}\,
   ((\HJQ_{1}^{\,2})_{2}^{2})_{4}\,;
   \end{array}  \]
\[ \begin{array}{ll}
   \lefteqn{\!\!J^P=3^-:}\\
&  [[\HSQ_{2},\HSQ_{1}]_{1},\HSQ_{2}]_{3} - 
   i\,{\frac{5}{3}}\,{\sqrt{5}}\,[[\HSQ_{2},\HSQ_{1}]_{3},\HSQ_{1}]_{3} + 
   i\,{\frac{43}{3}}\,{\sqrt{{\frac{1}{10}}}}\,
   [[\HSQ_{2},\HSQ_{1}]_{2},\HSQ_{1}]_{3} \\
&  - 12\,{\sqrt{{\frac{1}{5}}}}\,\{\HJQ_{1},[\HTQ_{2},\HSQ_{1}]_{3}\}_{3} - 
   {\frac{4}{3}}\,{\sqrt{10}}\,\{\HJQ_{1},[\HTQ_{2},\HSQ_{1}]_{2}\}_{3} - 
   i\,8\,{\sqrt{{\frac{3}{5}}}}\,\{\HTQ_{2},\HSQ_{2}\}_{3} 
   \qquad\qquad\qquad\qquad\qquad\qquad\\
&  + 16\,{\sqrt{{\frac{3}{5}}}}\,\{\HTQ_{2},\HSQ_{1}\}_{3} + 
   i\,{\frac{1264}{3}}\,{\sqrt{{\frac{1}{15}}}}\,
   \{(\HJQ_{1}^{\,2})_{2},\HSQ_{2}\}_{3} + 
   96\,{\sqrt{{\frac{3}{5}}}}\,\{(\HJQ_{1}^{\,2})_{2},\HSQ_{1}\}_{3}\,;
   \end{array}  \]
\[ \begin{array}{ll}
   \lefteqn{\!\!J^P=2^+:} \\
&  [[\HTQ_{2},\HSQ_{1}]_{1},\HSQ_{2}]_{2} - 
   {\frac{23}{20}}\,{\sqrt{{\frac{1}{35}}}}\,
   [[\HSQ_{2},\HSQ_{1}]_{2},\HTQ_{2}]_{2} - 
   {\frac{41}{60}}\,[[\HSQ_{2},\HSQ_{1}]_{1},\HTQ_{2}]_{2}
   \qquad\qquad\qquad\qquad\qquad\qquad\\
&  - i\,{\frac{61}{450}}\,{\sqrt{{\frac{1}{14}}}}\,
   [[\HTQ_{2},\HSQ_{1}]_{3},\HSQ_{1}]_{2} + 
   i\,{\frac{287}{180}}\,{\sqrt{{\frac{1}{10}}}}\,
   [[\HTQ_{2},\HSQ_{1}]_{2},\HSQ_{1}]_{2} + 
   i\,{\frac{13}{100}}\,{\sqrt{{\frac{1}{6}}}}\,
   [[\HTQ_{2},\HSQ_{1}]_{1},\HSQ_{1}]_{2} \\
&  + {\frac{116}{25}}\,{\sqrt{{\frac{2}{7}}}}\,
   \{\HJQ_{1},[\HSQ_{2},\HSQ_{1}]_{3}\}_{2} -
   {\frac{68}{15}}\,{\sqrt{{\frac{2}{5}}}}\,
   \{\HJQ_{1},[\HSQ_{2},\HSQ_{1}]_{2}\}_{2} + 
   {\frac{51}{25}}\,{\sqrt{{\frac{3}{2}}}}\,
   \{\HJQ_{1},[\HSQ_{2},\HSQ_{1}]_{1}\}_{2} \\
&  - {\frac{i}{2}}\,{\sqrt{{\frac{1}{5}}}}\,
   \{\HJQ_{1},[\HSQ_{1},\HSQ_{1}]_{1}\}_{2} - 
   i\,{\frac{436}{5}}\,{\sqrt{{\frac{2}{105}}}}\,(\HTQ_{2}^{\,2})_{2} - 
   i\,{\frac{171}{5}}\,{\sqrt{{\frac{6}{35}}}}\,(\HSQ_{2}^{\,2})_{2} \\
&  - {\frac{1}{2}}\,{\sqrt{{\frac{1}{15}}}}\,\{\HSQ_{2},\HSQ_{1}\}_{2} - 
   i\,{\frac{4412}{15}}\,{\sqrt{{\frac{2}{105}}}}\,
   \{(\HJQ_{1}^{\,2})_{2},\HTQ_{2}\}_{2} + 
   i\,{\frac{616}{15}}\,{\sqrt{{\frac{2}{15}}}}\,
   \{(\HJQ_{1}^{\,2})_{0},\HTQ_{2}\}_{2} \\
&  - i\,{\frac{1152}{35}}\,{\sqrt{{\frac{6}{5}}}}\,
   \{(\HJQ_{1}^{\,2})_{0},(\HJQ_{1}^{\,2})_{2}\}_{2} + 
   i\,({\frac{10492}{75}}-{\frac{109}{10}}\,f)\,
   {\sqrt{{\frac{1}{10}}}}\,\HTQ_{2}\\
&  - i\,({\frac{39168}{175}}+{\frac{78}{5}}\,f)\,
   {\sqrt{{\frac{2}{5}}}}\,(\HJQ_{1}^{\,2})_{2}\,; \vspace{2mm}\\
&  [[\HSQ_{2},\HSQ_{1}]_{3},\HTQ_{2}]_{2} - 
   {\sqrt{{\frac{2}{5}}}}\,[[\HSQ_{2},\HSQ_{1}]_{2},\HTQ_{2}]_{2} + 
   {\frac{1}{3}}\,{\sqrt{14}}\,[[\HSQ_{2},\HSQ_{1}]_{1},\HTQ_{2}]_{2} - 
   i\,{\frac{44}{45}}\,[[\HTQ_{2},\HSQ_{1}]_{3},\HSQ_{1}]_{2} \\
&  - {\frac{i}{9}}\,{\sqrt{{\frac{7}{5}}}}\,
   [[\HTQ_{2},\HSQ_{1}]_{2},\HSQ_{1}]_{2} + 
   {\frac{i}{5}}\,{\sqrt{{\frac{7}{3}}}}\,
   [[\HTQ_{2},\HSQ_{1}]_{1},\HSQ_{1}]_{2} - 
   {\frac{116}{45}}\,\{\HJQ_{1},[\HSQ_{2},\HSQ_{1}]_{3}\}_{2} \\
&  + {\frac{16}{9}}\,{\sqrt{{\frac{7}{5}}}}\,
   \{\HJQ_{1},[\HSQ_{2},\HSQ_{1}]_{2}\}_{2} + 
   {\frac{4}{5}}\,{\sqrt{{\frac{7}{3}}}}\,
   \{\HJQ_{1},[\HSQ_{2},\HSQ_{1}]_{1}\}_{2} + 
   i\,64\,{\sqrt{{\frac{1}{15}}}}\,(\HTQ_{2}^{\,2})_{2} \\
&  + i\,24\,{\sqrt{{\frac{3}{5}}}}\,(\HSQ_{2}^{\,2})_{2} - 
   {\frac{2}{3}}\,{\sqrt{{\frac{70}{3}}}}\,\{\HSQ_{2},\HSQ_{1}\}_{2} + 
   i\,{\frac{608}{3}}\,{\sqrt{{\frac{1}{15}}}}\,
   \{(\HJQ_{1}^{\,2})_{2},\HTQ_{2}\}_{2} \\
&  - i\,{\frac{64}{3}}\,{\sqrt{{\frac{7}{15}}}}\,
   \{(\HJQ_{1}^{\,2})_{0},\HTQ_{2}\}_{2} + 
   i\,128\,{\sqrt{{\frac{3}{35}}}}\,
   \{(\HJQ_{1}^{\,2})_{0},(\HJQ_{1}^{\,2})_{2}\}_{2} - 
   i\,({\frac{704}{15}}-4\,f)\,{\sqrt{{\frac{7}{5}}}}\,\HTQ_{2} \\
&  + i\,({\frac{5472}{5}}+84\,f)\,{\sqrt{{\frac{1}{35}}}}\,
   (\HJQ_{1}^{\,2})_{2}\,;
   \end{array}  \]
\[ \begin{array}{ll}
   \lefteqn{\!\!J^P=2^-:}\\
&  [[\HSQ_{2},\HSQ_{1}]_{1},\HSQ_{2}]_{2} - 
   i\,{\frac{11}{15}}\,{\sqrt{{\frac{7}{2}}}}\,
   [[\HSQ_{2},\HSQ_{1}]_{3},\HSQ_{1}]_{2} + 
   i\,{\frac{7}{6}}\,{\sqrt{{\frac{1}{10}}}}\,
   [[\HSQ_{2},\HSQ_{1}]_{2},\HSQ_{1}]_{2}
   \qquad\qquad\qquad\qquad\qquad\qquad\\
&  + {\frac{i}{10}}\,{\sqrt{{\frac{3}{2}}}}\,
   [[\HSQ_{2},\HSQ_{1}]_{1},\HSQ_{1}]_{2} - 
   i\,{\frac{3}{5}}\,\{\HJQ_{1},[\HTQ_{2},\HSQ_{2}]_{1}\}_{2} + 
   {\frac{1}{5}}\,{\sqrt{{\frac{7}{2}}}}\,
   \{\HJQ_{1},[\HTQ_{2},\HSQ_{1}]_{3}\}_{2} \\
&  - {\frac{7}{3}}\,{\sqrt{{\frac{1}{10}}}}\,
   \{\HJQ_{1},[\HTQ_{2},\HSQ_{1}]_{2}\}_{2} - 
   {\frac{4}{5}}\,{\sqrt{6}}\,\{\HJQ_{1},[\HTQ_{2},\HSQ_{1}]_{1}\}_{2} + 
   i\,{\sqrt{{\frac{42}{5}}}}\,\{\HTQ_{2},\HSQ_{2}\}_{2} \\
&  - i\,{\frac{172}{15}}\,{\sqrt{{\frac{14}{15}}}}\,
   \{(\HJQ_{1}^{\,2})_{2},\HSQ_{2}\}_{2} + 
   i\,{\frac{56}{15}}\,{\sqrt{{\frac{2}{15}}}}\,
   \{(\HJQ_{1}^{\,2})_{0},\HSQ_{2}\}_{2} + 
   48\,{\sqrt{{\frac{3}{5}}}}\,\{(\HJQ_{1}^{\,2})_{2},\HSQ_{1}\}_{2} \\
&  + i\,({\frac{92}{3}}-{\frac{21}{2}}\,f)\,
   {\sqrt{{\frac{1}{10}}}}\,\HSQ_{2}\,; \vspace{2mm}\\
&  [[\HSQ_{1},\HSQ_{1}]_{1},\HSQ_{1}]_{2} + 
   i\,24\,{\sqrt{{\frac{1}{5}}}}\,\{\HJQ_{1},[\HTQ_{2},\HSQ_{2}]_{1}\}_{2} - 
   2\,{\sqrt{{\frac{14}{5}}}}\,\{\HJQ_{1},[\HTQ_{2},\HSQ_{1}]_{3}\}_{2} \\
&  + 2\,{\sqrt{2}}\,\{\HJQ_{1},[\HTQ_{2},\HSQ_{1}]_{2}\}_{2} - 
   6\,{\sqrt{{\frac{6}{5}}}}\,\{\HJQ_{1},[\HTQ_{2},\HSQ_{1}]_{1}\}_{2} - 
   4\,{\sqrt{3}}\,\{\HTQ_{2},\HSQ_{1}\}_{2} \\
&  + i\,{\frac{32}{5}}\,{\sqrt{{\frac{14}{3}}}}\,
   \{(\HJQ_{1}^{\,2})_{2},\HSQ_{2}\}_{2} + 
   i\,{\frac{224}{5}}\,{\sqrt{{\frac{2}{3}}}}\,
   \{(\HJQ_{1}^{\,2})_{0},\HSQ_{2}\}_{2} + 
   40\,{\sqrt{3}}\,\{(\HJQ_{1}^{\,2})_{2},\HSQ_{1}\}_{2}
%   \vspace{1mm}\\ &
   - i \,\frac{16}{5}\, \sqrt{2}\,\HSQ_2 \,;
   \end{array}  \]
\[ \begin{array}{ll}
   \lefteqn{\!\!J^P=1^+:}\\
&  [[\HSQ_{2},\HSQ_{1}]_{3},\HTQ_{2}]_{1} - 
   {\frac{1}{2}}\,{\sqrt{35}}\,[[\HSQ_{2},\HSQ_{1}]_{2},\HTQ_{2}]_{1} + 
   {\frac{1}{2}}\,{\sqrt{21}}\,[[\HSQ_{2},\HSQ_{1}]_{1},\HTQ_{2}]_{1}
   \qquad\qquad\qquad\qquad\qquad\qquad \vspace{1mm}\\
&  - {\frac{i}{2}}\,{\sqrt{{\frac{105}{2}}}}\,
   [[\HTQ_{2},\HSQ_{1}]_{2},\HSQ_{1}]_{1} + 
   {\frac{i}{2}}\,{\sqrt{{\frac{35}{2}}}}\,
   [[\HTQ_{2},\HSQ_{1}]_{1},\HSQ_{1}]_{1} + 
   {\sqrt{210}}\,\{\HJQ_{1},[\HSQ_{2},\HSQ_{1}]_{2}\}_{1} \vspace{1mm}\\
&  + 3\,{\sqrt{35}}\,\{\HSQ_{2},\HSQ_{1}\}_{1} + 
   i\,8\,{\sqrt{210}}\,\{(\HJQ_{1}^{\,2})_{2},\HTQ_{2}\}_{1}\,;
   \end{array}  \]
\[ \begin{array}{ll}
   \lefteqn{\!\!J^P=1^-:}\\
&  [[\HSQ_{2},\HSQ_{1}]_{1},\HSQ_{2}]_{1} - 
   i\,{\frac{11}{4}}\,{\sqrt{{\frac{1}{10}}}}\,
   [[\HSQ_{2},\HSQ_{1}]_{2},\HSQ_{1}]_{1} + 
   i\,{\frac{5}{4}}\,{\sqrt{{\frac{5}{6}}}}\,
   [[\HSQ_{2},\HSQ_{1}]_{1},\HSQ_{1}]_{1} 
   \qquad\qquad\qquad\qquad\qquad\qquad\\
&  + {\frac{7}{24}}\,[[\HSQ_{1},\HSQ_{1}]_{1},\HSQ_{1}]_{1} - 
   i\,{\frac{3}{2}}\,{\sqrt{{\frac{1}{5}}}}\,
   \{\HJQ_{1},[\HTQ_{2},\HSQ_{2}]_{1}\}_{1} - 
   i\,6\,{\sqrt{{\frac{1}{5}}}}\,
   \{\HJQ_{1},[\HTQ_{2},\HSQ_{2}]_{0}\}_{1} \\
&  - {\frac{7}{3}}\,{\sqrt{10}}\,
   \{\HJQ_{1},[\HTQ_{2},\HSQ_{1}]_{2}\}_{1} + 
   {\frac{49}{2}}\,{\sqrt{{\frac{1}{30}}}}\,
   \{\HJQ_{1},[\HTQ_{2},\HSQ_{1}]_{1}\}_{1} + 
   i\,3\,{\sqrt{{\frac{1}{10}}}}\,\{\HTQ_{2},\HSQ_{2}\}_{1} \\
&  - {\frac{49}{2}}\,{\sqrt{{\frac{1}{15}}}}\,\{\HTQ_{2},\HSQ_{1}\}_{1} + 
   i\,{\frac{32}{3}}\,{\sqrt{10}}\,\{(\HJQ_{1}^{\,2})_{2},\HSQ_{2}\}_{1} - 
   {\frac{161}{3}}\,{\sqrt{{\frac{1}{15}}}}\,
   \{(\HJQ_{1}^{\,2})_{2},\HSQ_{1}\}_{1} \\
&  + {\frac{98}{3}}\,{\sqrt{{\frac{1}{3}}}}\,
   \{(\HJQ_{1}^{\,2})_{0},\HSQ_{1}\}_{1} + 
   ({\frac{98}{15}} - {\frac{49}{8}}\,f)\,\HSQ_{1}\,. 
   \end{array}  \]
\begin{center}  \underline{Quantum actions of $\BHQ_0^{(1)}$:}
\end{center}
\[ \begin{array}{l}
\!\![{\BHQ_{0}^{(1)}},\THQ_{2}]_{2} = i\,{\sqrt{6}}\,
    [\SHQ_{2},\SHQ_{1}]_{2}\,; \vspace{1mm}\\
\!\!\lbrack {\BHQ_{0}^{(1)}},\SHQ_{2}\rbrack_{2} =
    - i\,2\,{\sqrt{{\frac{2}{3}}}}\,[\THQ_{2},\SHQ_{1}]_{2}
    + 2\,{\sqrt{{\frac{2}{3}}}}\,\{\JHQ_{1},\SHQ_{2}\}_{2}
    + i\,6\,\{\JHQ_{1},\SHQ_{1}\}_{2}\,;\\
\!\!\lbrack \BHQ_{0}^{(1)},\SHQ_{1}\rbrack_{1} =
    - i\,6\,{\sqrt{{\frac{2}{5}}}}\,[\THQ_{2},\SHQ_{2}]_{1}
    + 2\,{\sqrt{{\frac{3}{5}}}}\,[\THQ_{2},\SHQ_{1}]_{1} 
%\\&&
    - i\,12\,{\sqrt{{\frac{3}{5}}}}\,\{\JHQ_{1},\SHQ_{2}\}_{1}
    - 6\,{\sqrt{2}}\,\{\JHQ_{1},\SHQ_{1}\}_{1}\,.
    \end{array}  \]
\underline{Remarks}:
\\[2mm]
{\it 1.}:  Each restriction on the deformation parameters originates
from one or more consistency relations with degree $(l+2)$ in
the form of a condition stating that the coefficient in front of a given
selected basis element of a certain subspace $\FH^{l^\p}_{j,\pm}$,
$l^\p \le l+1$, must vanish. This allows to assign a definite order 
in Planck's constant, {\it viz}.\ the order $(l+2-l^\p)\ge1$, to
each equation from a complete collection of parameter restrictions
when the latter are presented in their original form.
\\[2mm]
{\it 2.}:  The common numerical value zero determined above for all 
twenty-six parameters, initially of first order in Planck's constant,
and of one parameter, initially of third order in Planck's constant,
does not come as a surprise. Indeed, notice that the inhomogeneities
of the system of equations for the parameters arise exclusively
from rearrangements of double anticommutators or, more specifically,
from the contributions of the respective double commutators to
these rearrangements. Thus, provided the parameter restrictions
are presented in their original form, the inhomogeneities appear
only in parameter equations of {\it even} positive integer order
in Planck's constant and, in particular, not in the equations of
first and third order determining the aforementioned $26+1$
parameters.
\\[2mm]
{\it 3.}:  Rescaling, if necessary, the deformation parameters by
square roots of appropriate rational numbers, it is possible to
give all independent parameter equations derived in the course of
the cycles of degree $\le(l+1)$ simultaneously a polynomial form
with {\it rational} coefficients. Moreover, without loss of
generality, it may be assumed that a definite order in Planck's
constant is assigned to each parameter equation in such form.

\vspace{2mm}
Based on the experience gained by the cycles of degree two and three,
I shall risk a prognosis for the next cycle $(l+1)=4$:
\\[2mm]
{\it 1.}:  There is a good chance that the one residual free parameter
$f$ will be fixed at a rational value in the course of the next cycle 
(\ie\ $l+1=4$) by the consistency requirements for all $\UH$--relations
with degree 5, spin and parity $J^P = 4^+$ induced from $\UH$--relations
with degrees $\le 4$. Typically, each such non-trivial consistency
condition would give rise to 6 linear inhomogeneous equations for the
only indeterminate $f$.
\\
The consistency of all $\UH$--relations with degree 5, spin 
and parity $J^P = 6^\pm,\,5^\pm$ obtained by induction from
the $\UH$--relations (and the $\BHQ_0^{(l)}$--actions)
with degree $\le 4$ is already guaranteed by the consistency
of the classical algebra $\gh$: on the one hand the quantum
corrections for the $\UH$--generating relations with degree
$\le 4$ do not survive these inductions, and on the other hand
rearrangements of the various terms carrying spin and parity
$J^P = 6^\pm,\,5^\pm$ cannot produce inhomogeneities since
the subspaces $\UH^3_{6,\pm}$, $\UH^3_{5,\pm}$ consist of the
zero element only.
\\[2mm]
{\it 2.}: In the course of this next cycle ($l+1=4$), the classical
action of $\B_0^{(3)}$ on the generators of $\U$ is deformed into
the quantum action of $\BHQ_0^{(3)}$ at the expense of 47 + 9 + 4
real parameters, in their initial form of first, second and third
order in Planck's constant, respectively.
\\
At first sight this rapidly growing number of free parameters may
cause a shock. However, a closer look suggests that the number of
consistency restrictions on these parameters grows even more rapidly. 
As a matter of fact, $\BHQ_0^{(3)}$ contributes to the $\UH$--relations
with degree $l+2=5$ in two ways: in the first place it promotes the
$\UH$--relations with degree 2 to $\UH$--relations with degree 5
(one $J^P = 4^-$--, two $J^P = 3^+$--, one $J^P =3^-$--, one $J^P 
= 2^-$-- and one $J^P = 1^+$--relation) and in the second place,
by the postulated commutativity of $\BHQ_0^{(1)}$ and $\BHQ_0^{(3)}$,
it provides three more $\UH$--relations with degree 5 (for $J^P = 2^+,
\,2^-,\,1^-$ one relation each). But then, consistency requires among other 
things that the commutator action of $\BHQ_0^{(3)}$ annihilates the 
element $[\HTQ_{2},\HSQ_{2}]_{4}$. This condition by itself, confronted
with the rest of the $\UH$--relations with degree 5, spin and parity 
$4^-$, will already furnish 19 linear homogeneous and 4 quadratic
inhomogenous equations for the 38 real parameters involved: 32 of them
initially of order $\hslash^1$, and the remaining 6 parameters (including
the residual parameter $f$) initially of order $\hslash^2$.
\\[2mm]
{\it 3.}:  It is obvious from the order of $\BHQ_0^{(3)}$ in
Planck's constant that the 60 real indeterminates parametrizing
the quantum action of $\BHQ_0^{(3)}$ appear only linearly in
the parameter equations which turn up as consistency conditions
in the course of the deformation cycle of degree $(l+1)=4$. If
one way or another the numerical value of the parameter $f$ can
be determined as a rational number, the system of equations for
the remaining parameters will decouple into two subsystems: one
subsystem consisting of the homogeneous linear equations with
rational coefficients of first and third (possibly even fifth)
order in Planck's constant for the $47+4$ parameters, initially
of first and third order in Planck's constant, the other subsystem
consisting of the inhomogeneous linear equations with rational
coefficients of second and fourth order in Planck's constant for
the 9 parameters, initially of second order in Planck's constant.
\\
If the equations of the first subsystem combine to determine
some or all parameters involved, the values of these parameters
must be zero. If the analogous assumption applies to the second
subsystem, the values of the pertinent parameters will be rational
numbers. It is tempting to speculate that ultimately all parameters
will be determined, those involved in the first subsystem necessarily
as zero, those involved in the second subsystem necessarily as 
rational numbers.
\\
If the speculation is borne out by facts, then the vanishing
of the former parameters would suggest that in the quantum theory
a $\Zet_2$--grading survives as a reminiscence of the classical
$\N_0$--grading w.r.t.\ the degree $l$, whereas the rational values
of the latter parameters would match well with the circumstance
that the classical invariant charges form a Poisson algebra with
integer numbers as ``structure constants''.

\section*{Conclusions}

Exact quantum theoretic information in such detail as above
is available only for systems with an exceptionally large
symmetry group. In the context of physics the interest in
such a system is clearly justified if this system serves as
a model for some concrete physical phenomenon providing a
satisfactory description of the principal features of the
phenomenon. Apart from that, the interest in such a system
is also justified if as a model this system by itself leaves
much to be desired, but if -- as far as mathematical structure,
concepts and methods are concerned -- this system serves as a
good starting point for a systematic improvement and extension
procedure. As I see it, the interest in the Nambu--Goto theory
and in its structural analysis is of the second kind.

On account of the rapidly growing computational demands,
alternative ways of gathering information must be explored
in order to arrive at a purely algebraic description of the
most important observable features of the Nambu--Goto theory.
One such way is the clarification of the bialgebra aspects
of the quantum \aoo. The non-additive composition laws for
the invariant charges of two branches of a string trajectory
merging in a single branch, which have been established in
Ref.\ \cite{NonAd}, hint at the existence of a non-trivial
coproduct of the \aoo. The adequate setting for the
construction of such a coproduct is the minimal algebra
embedding the individual algebras corresponding to the three
separate branches. This embedding algebra contains also
elements which cannot be affiliated with any branch as
piecewise conserved charges, which must therefore be assigned to
the vertex itself.

Also for another reason this algebra is interesting: the
energy--momentum operators corresponding to the individual
merging branches cease to be central elements, and so do
their mass squares. This means that, apart from the operators
which were studied so far and which raise the degree, the
algebra contains also operators lowering the degree. With
their help additional diagonalizable elements with corresponding
spectra/roots can be produced beyond those which furnish
the quantum labels degree, spin and parity employed throughout.

It was pointed out to me by L.\ Tisza that the idea to formulate
the observable features of a continuum field theory in purely
algebraic terms and their physical interpretation was already
pondered by A.\ Einstein \cite{Ein}. This fact does not seem to
be a matter of common knowledge.
\\[5mm]
{\it Acknowledgements}:
I thank G.\ Handrich for clarifying discussions and for taking
his share in the tedious calculations. Computational assistance
by him as well as by T.\ Fischer, especially in the form of a final
computer check of the relations, is gratefully acknowledged.

I would like to thank the staff of the Physics Department of the
University of California at Berkeley for the kind hospitality
extended to me during the fall of 1996 and the spring of 1997.

This work was supported in part by Volkswagen--Stiftung and in part
by U.S.\ National Science Foundation under grant PHY--95--14797.

\end{document}